\documentclass[traditabstract]{aa}  
\usepackage{graphicx}
\usepackage{natbib}
\bibpunct{(}{)}{;}{a}{}{,} 
\usepackage[varg]{txfonts}
\usepackage{hyperref}
\newcommand{\mearth}{M_\oplus}

\newcommand{\rstar}{R_\star}

\newcommand{\rplanet}{R_\mathrm{p}}

\def\ms{\hbox{\,m\,s$^{-1}$}}         
\def\m2s2{\hbox{\,m$^{2}$\,s$^{-2}$}} 

\def\Mjup{\hbox{$\mathrm{M}_{\rm Jup}$}}
\def\Rjup{\hbox{$\mathrm{R}_{\rm Jup}$}}

\begin{document}

   \title{The GAPS Programme with HARPS-N@TNG}

   \subtitle{V. A comprehensive analysis of the XO-2 stellar and planetary systems \thanks{Based on observations made (\textit{i}) with the Italian Telescopio Nazionale Galileo (TNG), operated on the island of La Palma by the INAF - Fundacion Galileo Galilei (Spanish Observatory of Roque de los Muchachos of the IAC); (\textit{ii}) with the Copernico and Schmidt telescopes (INAF - Osservatorio Astrofisico di Padova, Asiago, Italy); (\textit{iii}) with the IAC-80 telescope at the Teide Observatory (Instituto de Astrof{\'\i}sica de Canarias, IAC); (\textit{iv}) at the Serra la Nave "M.G. Fracastoro" Astronomical Observatory (INAF - Osservatorio Astrofisico di Catania); (\textit{v}) at the Astronomical Observatory of the Autonomous Region of the Aosta Valley (OAVdA).
}}

\author{
M. Damasso     \inst{1,2},
K. Biazzo      \inst{3},
A.S. Bonomo    \inst{1},
S. Desidera    \inst{4},
A.F. Lanza     \inst{3},
V. Nascimbeni  \inst{4,5},
M. Esposito    \inst{6,7}, 
G. Scandariato \inst{3},
A. Sozzetti    \inst{1},
R. Cosentino   \inst{3,8},
R. Gratton     \inst{4},
L. Malavolta   \inst{5,9},
M. Rainer      \inst{10},
D. Gandolfi    \inst{3,11},
E. Poretti     \inst{10},
R. Zanmar Sanchez      \inst{3},
I. Ribas       \inst{12},
N. Santos      \inst{13,14,15},
L. Affer       \inst{16},
G. Andreuzzi   \inst{8}, 
M. Barbieri    \inst{4},
L. R. Bedin  \inst{4},
S. Benatti     \inst{4},
A. Bernagozzi      \inst{2},
E. Bertolini      \inst{2},
M. Bonavita    \inst{4},
F. Borsa       \inst{10},
L. Borsato		\inst{5},
W. Boschin     \inst{8},
P. Calcidese   \inst{2}, 
A. Carbognani  \inst{2},
D. Cenadelli    \inst{2},
J.M. Christille \inst{2,17},
R.U. Claudi    \inst{4},
E. Covino      \inst{18},
A. Cunial		\inst{5},
P. Giacobbe    \inst{1},
V. Granata		\inst{5},
A. Harutyunyan \inst{8},
M. G. Lattanzi \inst{1},
G. Leto        \inst{3},
M. Libralato		\inst{4,5},
G. Lodato      \inst{19}, 
V. Lorenzi	\inst{8},
L. Mancini     \inst{20},
A.F. Martinez Fiorenzano \inst{8},
F. Marzari     \inst{4,5}, 
S. Masiero     \inst{4,5}, 
G. Micela      \inst{16},
E. Molinari    \inst{8,21},
M. Molinaro   \inst{22},
U. Munari      \inst{4},
S. Murabito    \inst{6,7},
I. Pagano      \inst{3},
M. Pedani      \inst{8},
G. Piotto      \inst{4,5},
A. Rosenberg   \inst{6,7},
R. Silvotti    \inst{1},
J. Southworth  \inst{23}
}

\institute{
INAF -- Osservatorio Astrofisico di Torino, Via Osservatorio 20, I-10025, Pino Torinese, Italy.\\
E-mail: \texttt{damasso@oato.inaf.it}
\and Osservatorio Astronomico della Regione Autonoma Valle d'Aosta,  Fraz. Lignan 39, I-11020, Nus (Aosta), Italy
\and INAF -- Osservatorio Astrofisico di Catania, Via S.Sofia 78, I-95123, Catania, Italy
\and INAF -- Osservatorio Astronomico di Padova,  Vicolo dell'Osservatorio 5, I-35122, Padova, Italy
\and Dip. di Fisica e Astronomia Galileo Galilei -- Universit\`a di Padova, Vicolo dell'Osservatorio 2, I-35122, Padova, Italy
\and Instituto de Astrof{\'i}sica de Canarias, C/Via L{\'a}ctea S/N, E-38200 La Laguna, Tenerife, Spain
\and Departamento de Astrof{\'i}sica, Universidad de La Laguna,  E-38205 La Laguna, Tenerife, Spain
\and Fundaci\'on Galileo Galilei - INAF, Rambla Jos\'e Ana Fernandez P\'erez 7, E-38712 Bre\~na Baja, TF - Spain
\and Obs. Astronomique de l'Univ. de Geneve, 51 ch. des Maillettes -  Sauverny, CH-1290, Versoix, Switzerland
\and INAF -- Osservatorio Astronomico di Brera, Via E. Bianchi 46, I-23807 Merate (LC), Italy
\and Landessternwarte K\"onigstuhl, Zentrum f\"ur Astronomie der Universitat Heidelberg, K\"onigstuhl 12, D-69117 Heidelberg, Germany
\and Inst. de Ciencies de l'Espai (CSIC-IEEC), Campus UAB, Facultat de Ciencies, 08193 Bellaterra, Spain 
\and Instituto de Astrof\'isica e Ci\^encias do Espa\c{c}o, Universidade do Porto, CAUP, Rua das Estrelas, 4150-762 Porto, Portugal 
\and Centro de Astrof{\'\i}sica, Universidade do Porto, Rua das Estrelas,  4150-762 Porto, Portugal
\and Departamento de F{\'\i}sica e Astronomia, Faculdade de Ci\^encias,  Univ. do Porto, Rua do Campo Alegre, s/n, 4169-007 Porto,  Portugal
\and INAF -- Osservatorio Astronomico di Palermo, Piazza del Parlamento, 1, I-90134, Palermo, Italy
\and Dept. of Physics, University of Perugia, via A. Pascoli, 06123, Perugia, Italy
\and INAF -- Osservatorio Astronomico di Capodimonte, Salita Moiariello 16, I-80131, Napoli, Italy
\and Dipartimento di Fisica, Universit\`a  di Milano, Via Celoria 16, I-20133 Milano, Italy
\and Max-Planck-Institut f\"ur Astronomie, K\"onigstuhl 17, D-69117, Heidelberg, Germany
\and INAF - IASF Milano, via Bassini 15, I-20133 Milano, Italy
\and INAF -- Osservatorio Astronomico di Trieste, via Tiepolo 11, I-34143 Trieste, Italy
\and Astrophysics Group, Keele University, Staffordshire, ST5 5BG, UK
 }

   \date{}

 
  \abstract
   {}
   {XO-2 is the first confirmed wide stellar binary system where the almost twin components XO-2N and XO-2S have planets, and it is a peculiar laboratory to investigate the diversity of planetary systems. This stimulated a detailed characterization study of the stellar and planetary components based on new observations.}
   {We collected high-resolution spectra with the HARPS-N spectrograph and multi-band light curves. Spectral analysis led to an accurate determination of the stellar atmospheric parameters and characterization of the stellar activity, and high-precision radial velocities of XO-2N were measured. We collected fourteen transit light curves of XO-2Nb used to improve the transit parameters. Photometry provided accurate magnitude differences between the stars and a measure of their rotation periods.}
   {The iron abundance of XO-2N was found +0.054 dex greater, within more than 3$\sigma$, than that of XO-2S. The existence of a long-term variation in the radial velocities of XO-2N is confirmed, and we detected a turn-over with respect to previous measurements. We suggest the presence of a second massive companion in an outer orbit or the stellar activity cycle as possible causes of the observed acceleration. The latter explanation seems more plausible with the present dataset. We obtained an accurate value of the projected spin-orbit angle for the XO-2N system (\textit{$\lambda$}=7$^{\circ}\pm$11$^{\circ}$), and estimated the real 3-D spin-orbit angle (\textit{$\psi$}=27$^{+12}_{-27}$ degrees). We measured the XO-2 rotation periods, and found a value of P=41.6$\pm$1.1 days in the case of XO-2N, in excellent agreement with the predictions. The period of XO-2S appears shorter, with an ambiguity between 26 and 34.5 days that we cannot solve with the present dataset alone. The analysis of the stellar activity shows that XO-2N appears to be more active than the companion, and this could be due to the fact that we sampled different phases of their activity cycle, or to an interaction between XO-2N and its hot Jupiter that we could not confirm.} 
  {}


   \authorrunning{M. Damasso et al.}
   \titlerunning{GAPS V: Global analysis of the XO-2 system}

   \maketitle
%

\section{Introduction}
The diversity of the more than 1,700 extrasolar planets discovered so far\footnote{NASA exoplanet archive, http://exoplanetarchive.ipac.caltech.edu/} ($\sim$850 of which are in multi-planet systems; \citealt{rowe14}), from their orbital architectures to the astrophysical environments where they reside, represents a very complex issue to be investigated to understand the mechanisms of their formation and dynamical evolution. 

	In the era of comparative exo-planetology, a fundamental approach for the understanding of the different exoplanetary properties relies on the characterization of planet-host stars and statistical analysis of the planetary system frequency. Several studies shown that the properties of the planetary system architectures and physical characteristics of the planets depend upon stellar properties, such as the mass \citep{johnson10, bonfils13} and the metallicity \citep{sozzetti09, santos11, mortier12}. Moreover, during the last years the search for exoplanets was extended in environments like giant stars (e.g. Kepler-91, \citealt{lillobox14a,lillobox14b}),open and globular clusters, and only recently planetary companions were found (see, e.g., \citealt{quinn12,meibom13}, and references therein).

	Binary systems are interesting astrophysical environments to search for planets. Considering that nearly half of the solar-type stars are gravitationally bound with at least another stellar companion (e.g. \citealt{raghavan10,duchene13}), binary systems naturally represent a typical environment to be explored for studying the processes of planet formation and evolution leading to very different planetary architectures. 
	
	Planet formation in a binaries has been considered by various authors. For coplanar orbits between the disk and the binary the general expectation (e.g., \citealt{marzari00, kley08, marzari12}) is that the tidal effect of the companion would induce an eccentricity growth in the planetesimal population that might inhibit planet formation if the binary separation is $\lesssim$ 50 AU. In this configuration even the formation of icy planetesimals is jeopardized by the high temperatures in the discs and the formation of breaking waves \citep{nelson00,picogna13}. The situation is more complicated for misaligned systems where the planetesimals and gas discs precess around the binary orbital plane. Planetesimals naturally develop strong differential precession that would inhibit planet formation by increasing the velocity dispersion \citep{marzari09}. On the other hand, if the gas disk is radially narrow (which is expected for separation of the order of 50 AU because of tidal truncation), it will precess rigidly and would drag the planetesimals into rigid precession, mitigating the previous effect \citep{fragner11}, unless the inclination is so large for Lidov-Kozai effects to become important. Also in this case, the effects should be relevant mostly for binary separations smaller than 50-100 AU. For wider, misaligned binaries the gas disk on the other hand is expected to reach a quasi-steady, non precessing warped configuration (see \citealt{facchini13} for the case of a circumbinary disk). If the warp is located in the planet formation region, it might affect the process, although such effects have not yet been studied. 	
	
	Several examples of planets orbiting only one of the binary components, defined as \textit{S-type} planets, are known (e.g. \citealt{roell12,mugrauer14}, and \citealt{thebault14} for a recent review). The existence of \textit{S-type} planets rises several questions: how the presence of a stellar companion can affect the formation, survival and dynamical evolution of such planets? Is it possible to identify which properties of a stellar system most influence the physical characteristics of the hosted planets? Does any constraint exist with respect to the case of isolated stars, or any dependence from the physical separation of the binary components? 
Observational difficulties exist in searching for \textit{S-type} planets in multiple stellar systems \citep{eggenberger10}, in particular those with a very small angular separation (less then $\sim$2$^{\prime\prime}$), since, for instance, the components cannot be observed as two isolated targets in the spectrograph’s fiber or slit. In many cases of confirmed extrasolar planets the presence of a stellar companion to the host star was discovered after the detection of the planet (e.g. \citealt{roell12}). Binarity poses a major challenge in understanding how such planets could exist and their evolutionary history in the stellar system.

	Very wide binaries, with semi-major axes of the order of 10$^{3}$ AU and sky projected angular separations of several arcsec, offer the best opportunity and less observational complications to search for planets orbiting each component. At the same time they allow an accurate determination of the properties of the individual stars. Despite the large separation between the stellar components, these systems are an interesting subject for dynamical studies also related to the presence of planets. 
	
\citet{kaib13} and \citet{kaib14} studied the perturbations produced by passing stars and the tidal effects of the Milky Way on the orbits of very wide binary stars. Interestingly they found that occasionally the orbits could become extremely eccentric, resulting in the collision of the binary components or, less dramatically, they could evolve toward close or contact systems. Moreover, simulations by \cite{kaib13} indicate that very wide binary companions may often strongly reshape planetary systems after their formation, by causing the ejection of planets from the system or heavily changing the orbital eccentricities of those that survive. Another intriguing example of a plausible fate for planets in wide binaries is the 'bouncing' scenario investigated by \cite{moeckel12}. They found that in 45 to 75 per cent of the cases a planet, initially orbiting one of the binary components and then ejected from its native system by planet-planet scattering, could jump up and down passing from the gravitational influence of one star to the other, for more than a Myr. This situation may trigger orbital instability among existing planets around the companion, and in some cases result in an exchange of planets between the two stars, when both host multiple planetary systems.
 
 	At present, few very wide binaries are known where a planetary system was discovered around one of the stellar components. The star HD20782 is the binary companion of HD20781, with a projected separation of $\sim$9000 AU, and it hosts a Jupiter-mass planet on a very eccentric orbit at $\sim$1.4 AU \citep{jones06}\footnote{HD20781 should host two Neptune-mass planets within $\sim$0.3 AU, as reported by \cite{mayor11}, but the discovery has not yet been confirmed and published.}. The very high eccentricity reported in the literature (\textit{e} = 0.97) could be the result of significant perturbations experienced by the planet during the evolution of the system and possibly due to the stellar companion. In this binary system, where both the components can be well analysed separately, a comparative study of the physical properties of the pair could be of help for explaining the nature of the planetary system. To investigate any existing link between the planet formation mechanisms and the chemical composition of the host stars, \cite{mack14} performed a detailed elemental abundance analysis of the atmospheres of HD20781/82, by considering both stars to host planets. This kind of study, when at least one component hosts a close-in giant planet, could result in the evidence of chemical imprints left in the parent star suggesting the ingestion of material from the circumstellar disk (and possibly also of planetary origin) driven by the dynamical evolution of the planet orbit. In fact, when a star with a close-in giant planet is found to be enriched with elements of high condensation temperature, as suggested by \cite{schuler11}, this can be related to the inward migration of the planet from the outer regions of the circumstellar disk, where it formed, to the present position closer to the star \citep{ida08, raymond11}. \cite{mack14} found for both stars a quite significant positive trends with the condensation temperature among the elemental abundances, and suggest that the host stars accreted rocky bodies with mass between 10 and 20 $\mearth$ initially formed interior to the location of the detected planets. Searches for abundance anomalies possibly caused by the ingestion of planetary material by the central star were also conducted by \cite{desidera04,desidera07} (and references therein) for a sample of wide binaries. These authors found that the amount of iron accreted by the nominally metal richer companion (in binaries with components having T$_{\rm eff}$>5500 K) is comparable to the estimates of the rocky material accreted by the Sun during its main-sequence lifetime, and therefore concluded that the metal enrichment due to the ingestion of material of planetary origin should not be a common event.
 
\begin{figure}
   \centering
   \includegraphics[width=9cm]{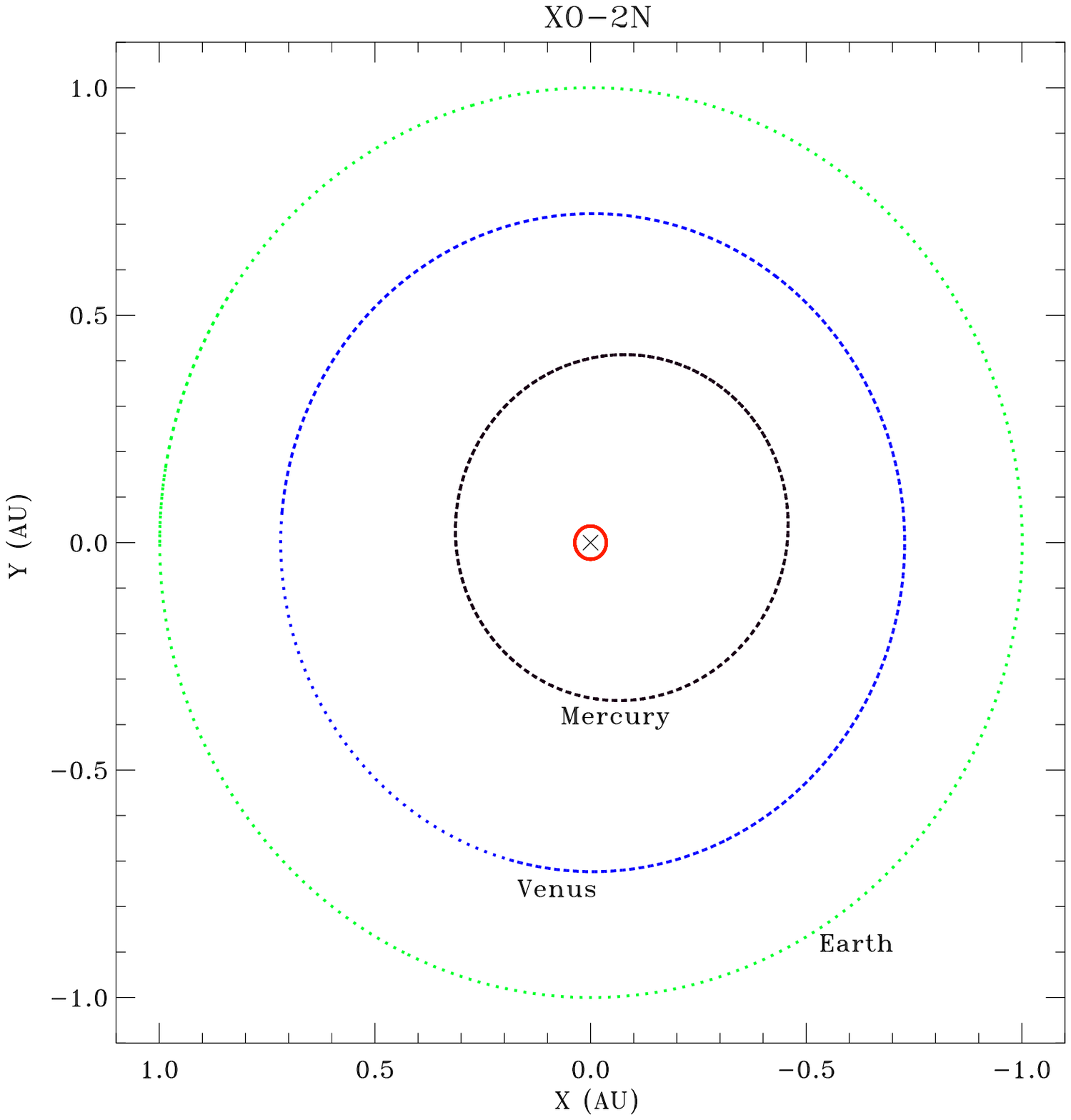}
   \includegraphics[width=9cm]{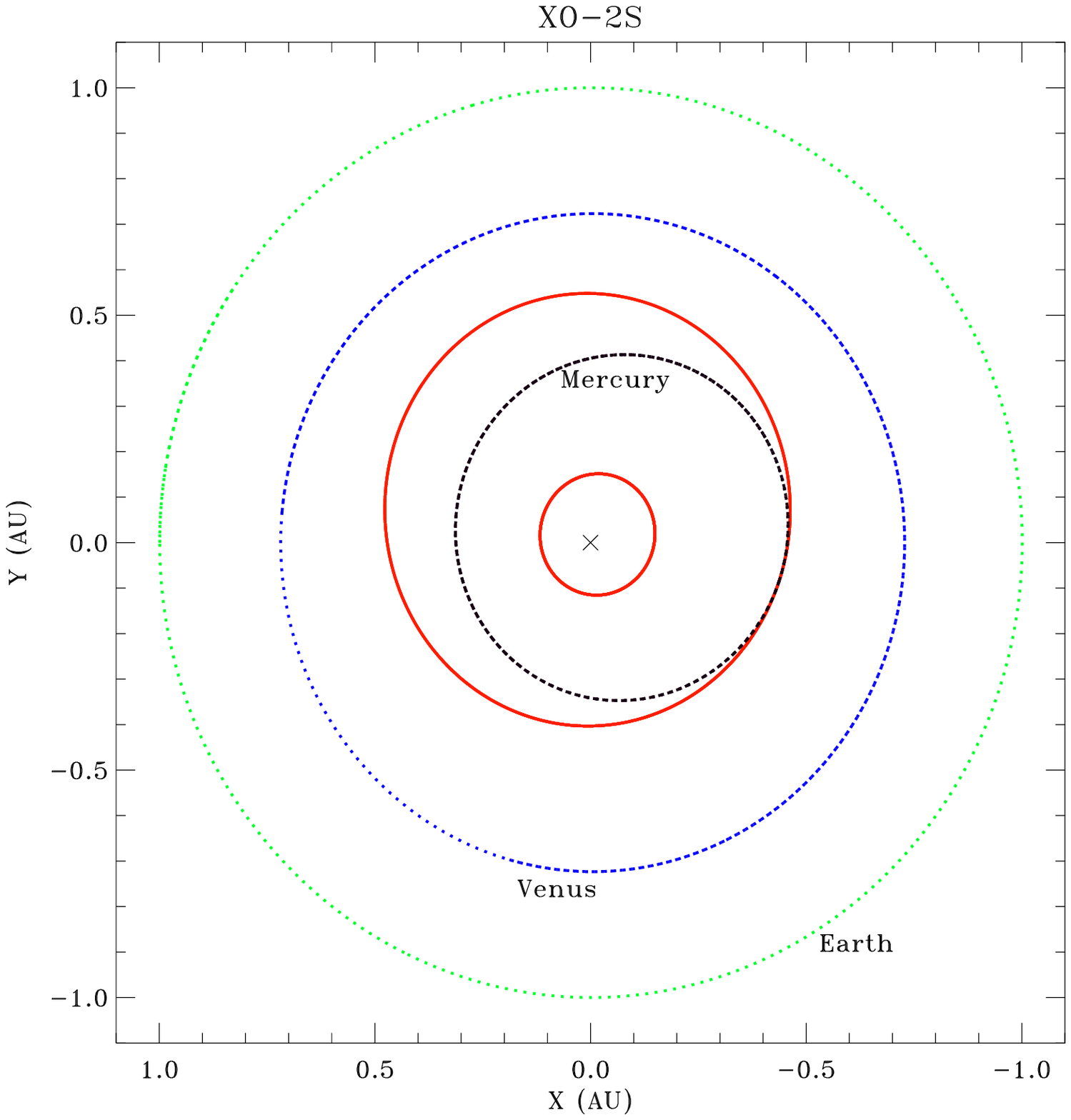}
   \caption{Comparison between the planetary systems orbiting the XO-2 binary stars and the architecture of the inner Solar System. The case of XO-2N is shown in the upper panel, while that of XO-2S is represented in the lower one. In both plots, the red ellipses (solid lines) represent the orbits of the XO-2 planets. The orbits of Mercury, Venus and the Earth are indicated with black, blue and green dotted lines respectively, also labelled with the planet names. The cross symbol in the middle of each plot indicates the star location.}
   \label{fig:planetorbits}%
    \end{figure}

 	A representative case of a very wide binary with one component hosting a close-in giant planet is that of the XO-2 system (projected separation $\sim$31$^{\prime\prime}$), where the star XO-2N is orbited by the transiting hot Jupiter XO-2Nb \citep{burke07}. An elemental abundance analysis of XO-2N and its companion XO-2S was performed by \citet{teske13}, who determined the stellar abundances of carbon and oxygen, as well as iron and nickel. Their goal was to probe the potential effects that planet formation and evolution might have had on the chemical composition of XO-2N, where the companion XO-2S was treated as a non-hosting planet star. 
 	Later, \cite{desidera14} discovered that the star XO-2S actually hosts a planetary system formed by a planet slightly more massive than Jupiter orbiting at 0.48 AU and a Saturn-mass planet at 0.13 AU (with cautious evidence of a long-term trend in the radial velocities time series possibly due to an outer companion, yet of unknown nature). The discovery is of particular relevance because it represents the first confirmed case of a very wide binary in which both components host planets (see Fig. \ref{fig:planetorbits} for a sketch of the XO-2 planetary systems). The XO-2 binary offers a unique opportunity to explore the diversity of planet formation mechanisms, by considering that, while the parent stars are almost equal in their main physical properties, the planetary systems are different. This finding has consequently motivated a detailed comparative study of the whole system, which is the subject of this paper, through the use of new high-resolution and high S/N HARPS-N spectra and dedicated differential photometry. It is crucial to highlight any existing difference between the properties of the two stars which could have played a distinctive role in the formation processes of the planetary systems: any observed difference, if correctly interpreted, can tell a part of that story. 
 	While the analysis of the XO-2 system was still in progress, the discovery of a hot Jupiter around each of the twins, metal-rich ([Fe/H]$\sim$0.25 dex), F-type components of the WASP-94 wide binary (projected separation $\sim$2700 AU) was announced \citep{neveu14}. This system, with host star properties and planetary system architectures different from those of XO-2, indeed makes more intriguing the understanding of the role of stellar multiplicity on the planet formation and evolution.  
 	
This paper is organized as follows. We first describe in Sect.\ref{sec:datared} the spectroscopic and photometric datasets used in this study. We present in Sect.\ref{sec:stellarpar} a new determination of the basic stellar parameters based on HARPS-N spectra, as well as results of a detailed and homogeneous analysis aimed at searching for differences in the iron abundance between the XO-2 components with high level of confidence. In Sect.\ref{sec:lcanalysis} we present and discuss results from a photometric follow-up of the two stars, including the measurement of the stellar rotation periods and the modelling of a new dataset of 14 transits of the planet XO-2Nb. Sect.\ref{sec:radvelanalysis} is dedicated to the XO-2N planetary system. There we present a new analysis of the radial velocity time series, thanks to which we confirm the existence of an acceleration with a not yet established cause. In the same Section we also present new observation and analysis of the Rossiter-McLaughlin effect, better constraining the value of the projected spin-orbit angle $\lambda$. A detailed analysis of the stellar activity for the XO-2 components is presented and discussed in Sect.\ref{sec:stellaractivity}. We finally conclude with Sect.\ref{sec:sysstab}, where we speculate about the evolution and long-term stability of the XO-2 planetary systems.

%


\section{Observations and data reduction methods}
\label{sec:datared}

	\subsection{Spectroscopy}
	\label{sec:datared1}
	The spectra of the XO-2 components analysed in this work were collected with the high-resolution HARPS-N spectrograph \citep{cosentino12} installed at the Telescopio Nazionale Galileo (TNG) on La Palma (Canary islands). The observations were carried out in the framework of the large programme Global Architecture of Planetary Systems (GAPS; \citealt{covino13,desidera13}).
	
	For the star XO-2N we collected 43 spectra between November 20, 2012 and October 4, 2014, while the companion XO-2S was observed at 63 individual epochs between April 21, 2013 and May 10, 2014. The Th-Ar simultaneous calibration was not used to avoid contamination by the lamp lines. This has no significant impact on the measurements of the radial velocity, because the drift correction with respect to the reference calibration shows a dispersion of 0.8 m s$^{-1}$, which is smaller than the median radial velocity (RV) uncertainty due to photon-noise, i.e. 2.0 m s$^{-1}$ for XO-2N and 2.2 m s$^{-1}$ for XO-2S.
	The reduction of the spectra and the RV measurements were obtained for both components using the latest version (Nov. 2013) of the HARPS-N instrument Data Reduction Software pipeline and applying a K5 mask. The measurement of the RVs is based on the weighted cross-correlation function (CCF) method \citep{baranne96,pepe02}.

	\subsection{Photometry}
	\label{sec:datared2}
     We collected and analysed new photometric light curves for both components of the XO-2 system with three different facilities. In particular, XO-2S was intensively monitored with the HARPS-N spectrograph. When it became clear that the star had planetary companions, we started a dedicated photometric follow-up of the target to estimate the stellar rotational period, further characterizing its level of activity and look for possible transits of its exoplanets.
    
    \subsubsection{The APACHE dataset}
	\label{sec:apachephotometry}
     A follow-up of the XO-2 field was conducted for 42 nights between December 2, 2013 and April 8, 2014 with one of the 40cm telescopes composing the APACHE array \citep{sozzetti13}, based at the Astronomical Observatory of the Autonomous Region of the Aosta Valley (OAVdA, +45.7895 N, +7.478 E, 1650 m a.s.l.). Each telescope is a Carbon Truss f/8.4 Ritchey-Chr$\'{e}$tien equipped with a GM2000 10-MICRON mount and a FLI Proline PL1001E-2 CCD Camera, with a pixel scale of 1.5 $^{\prime\prime}$/pixel and a field of view of 26$^{\prime}$x26$^{\prime}$. The observations were carried out using a Johnson-Cousins \textit{I} filter. The images were reduced with the standard pipeline TEEPEE written in \texttt{IDL}\footnote{Registered trademark of Exelis Visual Information Solutions.} and regularly used for the APACHE purposes (see \citealt{giacobbe12}). The APACHE data were used to tentatively derive the rotation period of the two stars and to monitor their magnitude difference in \textit{I} band, as described in Sect.\ref{sec:lcanalysis}.
    \subsubsection{The TASTE dataset}
	\label{sec:tastephotometry}
     Since February 2011 several transits of the hot Jupiter XO-2Nb have been observed at high temporal cadence with the 182cm and the Schmidt 92/67 telescopes at the Asiago Astrophysical Observatory (+45.8433 N, +11.5708 E, 1366m a.s.l.), in the framework of the TASTE Project (The Asiago Search for Transit time variations of Exoplanets; \citealt{nascimbeni11}). A subset of these measurements was used to estimate the magnitude differences of the two XO-2 binary components in the $R_c$ band. Two additional transits were also collected with the IAC-80 telescope at the Teide Observatory, using the CAMELOT CCD imager (see \citealt{nascimbeni13} for a description of the set-up), and one supplementary light curve was obtained with the same telescope but using the TCP camera (Troms\"{o} CCD Photometer; \citealt{ostensen00}). All these observations, for a total of 14 light curves, were carried out in the Cousin/Bessell $R$ band and were analysed together to improve the determination of the parameters of XO-2Nb. 
    \subsubsection{The Serra la Nave dataset}
	\label{sec:slnphotometry}
	 The XO-2 system was also observed for five nights, between May 6, 2013 and May 7, 2014 with the robotic 80cm f/8 Ritchey-Chr\'{e}tien telescope APT2, operated by the INAF-Catania Astrophysical Observatory and located at Serra la Nave (SLN, +37.692 N, +14.973 E, 1725m a.s.l.). The telescope is equipped with a CCD detector Apogee Alta U9000 (3Kx3K pixels) operated in binning 2x2, corresponding to a pixel scale 0.76 arcsec/pixel. The data were reduced with the \texttt{IRAF} package\footnote{http://iraf.noao.edu/} using the standard procedure for overscan, bias and dark subtraction and flat fielding. The aperture photometry was done with the software \texttt{SExtractor} \citep{bertin96}. The data of SLN were collected in the \textit{BVRI} bands and used to measure the magnitude differences between the two stars.


\section{Stellar parameters}
\label{sec:stellarpar}

	\subsection{Spectroscopic analysis of the individual stars}
	\label{sec:stellarpar1}
The components of the XO-2 binary system have a separation of $\sim$31 arcsec (corresponding to a projected distance of $\sim$4600 AU assuming a distance of $\sim$150 pc, as estimated by \citealt{burke07}), and share, within the uncertainties, the proper-motion vector as listed in the UCAC4 catalogue. Essential information about the XO-2 system is presented in Table \ref{table:stellarinfo}. 
A standard analysis of the HARPS-N spectra was performed with different methods to derive the stellar atmospheric parameters of the two components, with those for XO-2S first presented in \cite{desidera14}. Table \ref{table:stellarparam} summarizes our results, which represent the weighted averages of the individual measurements. We used implementations of both the equivalent width and the spectral synthesis methods, as described in \cite{biazzo12}, \cite{santos13}, and \cite{esposito14}. 
It is interesting to note that the effective temperature difference between the two stars derived spectroscopically is in good agreement with the result obtained from photometric measurements. In fact, following the same analysis performed by \cite{munari14} for half-million of stars of the RAVE spectroscopic survey, if we assume a color excess $E(B-V)$=0.02 mag (as explained in Sect.\ref{sec:stellarpar3}) the XO-2S star results 60$\pm$33 K hotter than the companion.
The stellar mass, radius and age were determined by comparing our measured effective temperature, iron abundance, and surface gravity with the Yonsei-Yale (Y-Y) evolutionary tracks \citep{demarque04} through the $\chi$-square statistics \citep{santerne11}. The adopted errors include an extra 5$\%$ in mass and 3$\%$ in radius added in quadrature to the formal errors to take systematic uncertainties in stellar models into account \citep{southworth11}. Taking advantage of the fact that the planet XO-2Nb transits its parent star, we derived a second, slightly more accurate estimate for the mass, radius and age of XO-2N, using the stellar density information as determined from the analysis of the transit light curves (e.g., \citealt{sozzetti07}).
    
    \subsection{Differential spectral analysis}
	\label{sec:stellarpar2}
	
\begin{figure*}[h!]
\includegraphics[width=9cm]{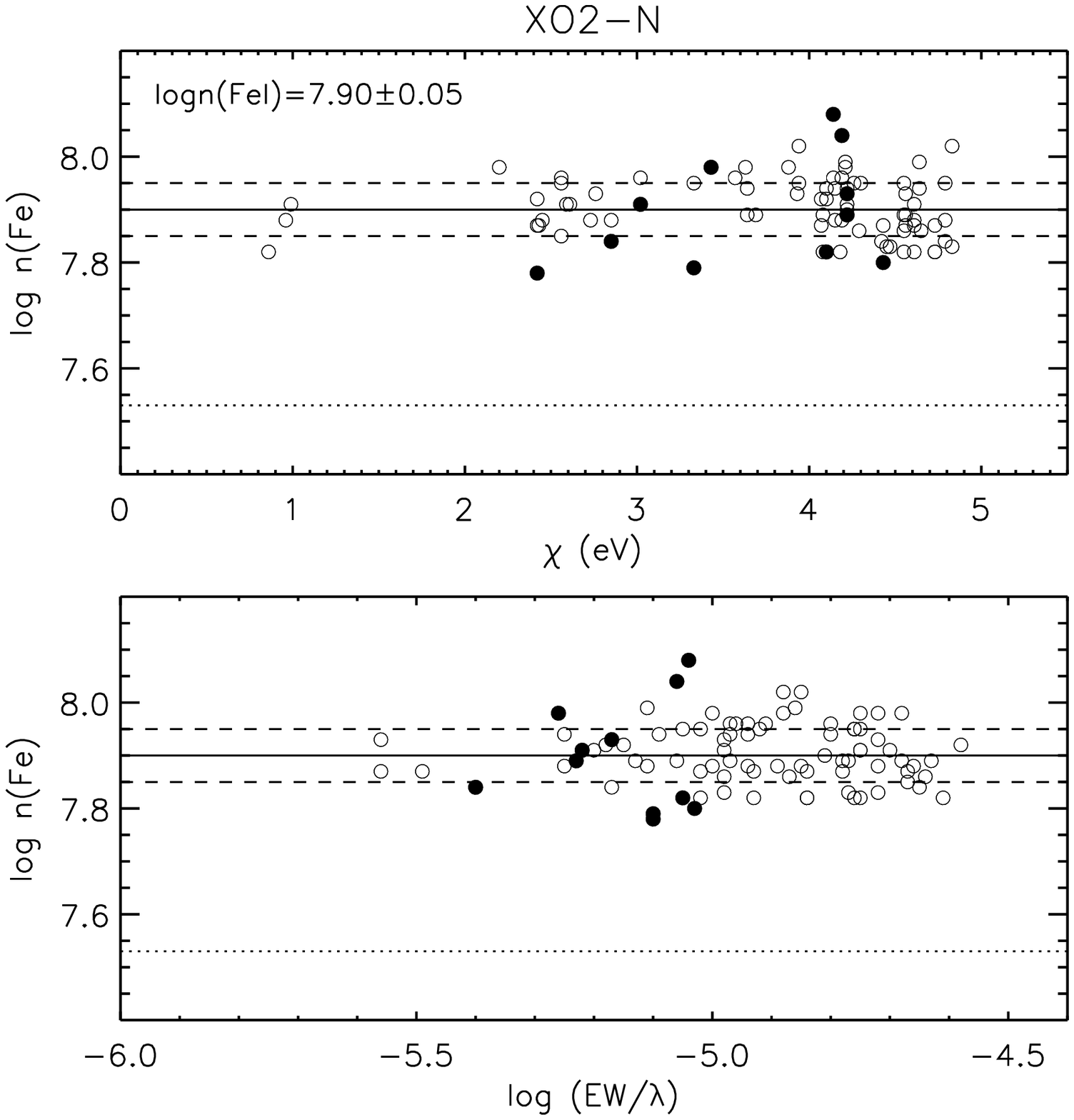}
\includegraphics[width=9cm]{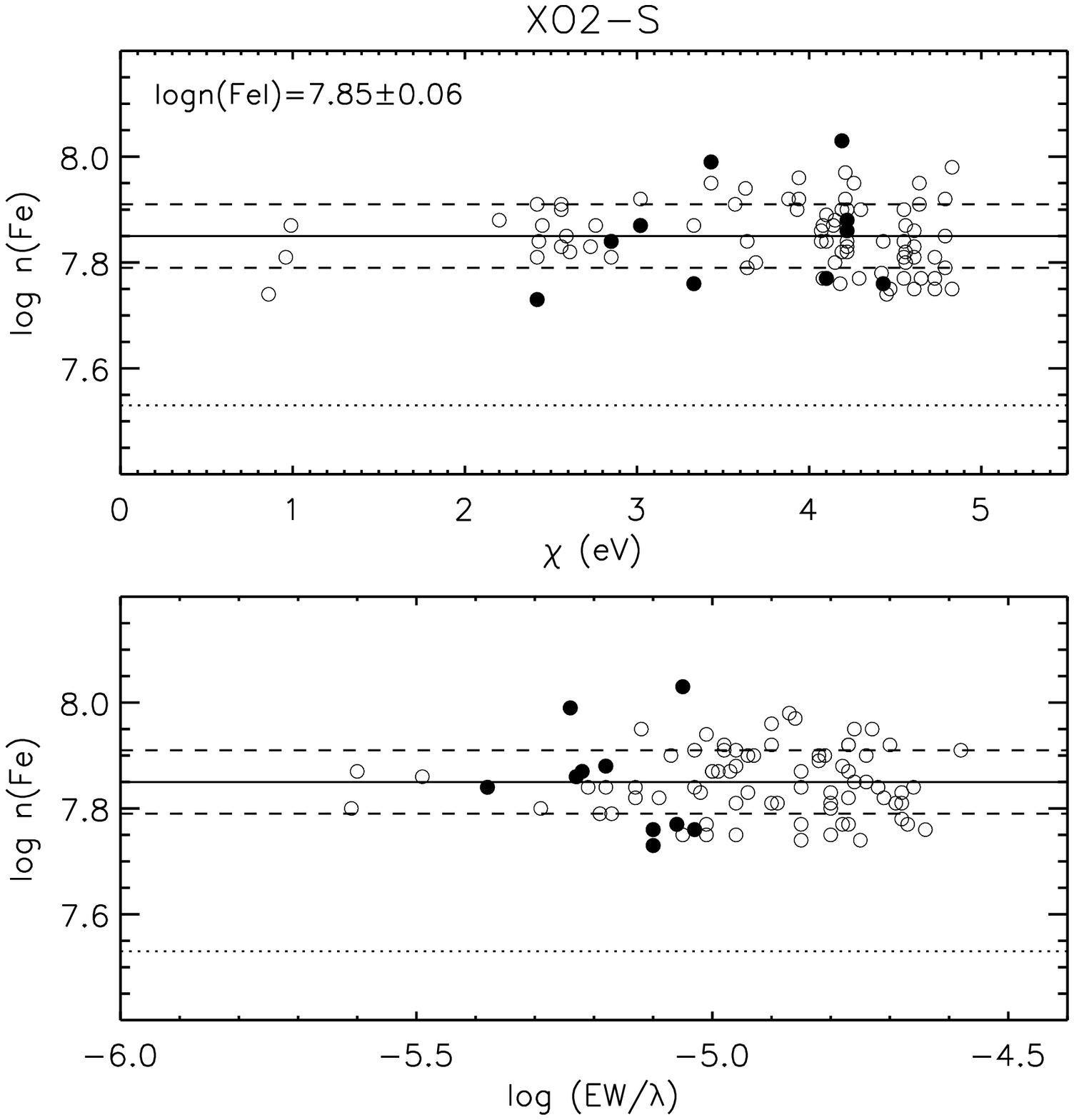}
\vspace{-.1cm}
\caption{Iron abundance of the XO-2 stellar binary derived through the analysis based on equivalent widths as 
a function of the excitation potential and reduced EW. Empty and filled circles represent the \ion{Fe}{i} and 
\ion{Fe}{ii} lines, respectively. Solid lines show the location of mean $\log n (\ion{Fe}{i})$, while dashed lines 
represent the standard deviation $\pm 1 \sigma$. The dotted line is intended to guide the eye to the solar iron abundance at 
$\log n{\rm (\ion{Fe}{i})_\odot}=7.53$.}
\label{fig:abund_csi_ew}
\end{figure*}

Thanks to the high-quality of the HARPS-N spectra, we performed a detailed and homogeneous spectral analysis of XO-2N and XO-2S to search for possible differences in the iron abundance between the components. The average values for [Fe/H] obtained from the analysis of the individual stars (Table \ref{table:stellarparam}) indicate that the North component is characterised by higher iron abundance, but the two estimates are compatible within the uncertainties. In order to validate this difference with a high level of confidence, we determined the difference in [Fe/H] very accurately with a dedicated analysis. Together with the spectra of both targets, we also acquired three solar spectra through observations of the asteroid Vesta. This allowed us to perform a differential analysis for each component with respect to the Sun, thus avoiding the contribution due to the uncertainties in atomic parameters, such as the transition probabilities. All HARPS-N spectra were shifted in wavelength and then co-added in order to obtain very high $S/N$ spectra ($\sim 180-250$ at $\lambda \sim 6700$ \AA). We performed the spectral analysis following three main steps:

{\it i. Analysis based on equivalent widths.} We measured the equivalent widths (EWs) of iron lines on one-dimensional 
spectra using the {\it splot} task in IRAF, paying attention to trace as much as possible the same position in the continuum 
level for the spectra of both components and the asteroid Vesta. We considered the line-list from \cite{biazzo12}, and 
adopted the same procedure as these authors for the differential analysis relative to the Sun, the rejection criteria of 
bad lines, the line broadening mechanisms, the model atmospheres, the measurement of stellar parameters and iron abundances, 
and their corresponding uncertainties. We refer to that paper and to \cite{biazzoetal2011} for detailed descriptions of the method. 
In brief, we used the version 2013 of the MOOG code (\citealt{sneden1973}) and considered the {\it abfind} driver, assuming local thermodynamic equilibrium (LTE). Initial stellar parameters were set to the solar values (T$_{\rm eff \odot}$=5770 K, $\log g_\odot=4.44$, and $\xi_\odot=1.10$ km/s) for both components. Then, the final effective temperature (T$_{\rm eff}$) was determined by imposing the condition that the abundance from the \ion{Fe}{i} lines ($\log n (\ion{Fe}{i})$) was independent on the line excitation potentials; the final microturbulence ($\xi$) by minimizing the slope of $\log n (\ion{Fe}{i})$ versus the reduced equivalent width ($EW/\lambda$); and the final surface gravity ($\log g$) by imposing ionization equilibrium (i.e. $\log n (\ion{Fe}{i})$=$\log n (\ion{Fe}{ii})$). Plots of iron abundance ($\log n {\rm (Fe)}$) versus excitation potential ($\chi$) and reduced EW for both components are shown in Fig.~\ref{fig:abund_csi_ew}, which shows that the correlations are close to zero, as required. Using the same procedure, we obtained $\log n{\rm (\ion{Fe}{i})_\odot}=\log n{\rm (\ion{Fe}{ii})_\odot}=7.53\pm0.05$ for the Sun. We point out that the exact values of the solar parameters are not crucial as we are performing a differential study for both components with respect to the Sun.
Table~\ref{tab:atmospheric_parameters} lists the final results of this procedure, where the errors in iron abundance include 
uncertainties in stellar parameters and in EW measurements. Results from this first homogeneous analysis seems to be in accordance with those obtained with different methods of spectral analysis (Table \ref{table:stellarparam}).

{\it ii. Line-by-line analysis.} Since we are interested in a homogeneous analysis, we applied the same iterative method as in the first step, but here we considered a strict line-by-line analysis, i.e. the same lines were used for both components, without any implementation of line rejection criteria. This allowed us to avoid uncertainties related to both atomic parameters and measurements of equivalent widths, due to, e.g., blends and continuum level. In the end, the analysis was based on 75 \ion{Fe}{i} and 10 \ion{Fe}{ii} lines. Thanks to such analysis, the errors in T$_{\rm eff}$ were reduced substantially, as reported in Table~\ref{tab:atmospheric_parameters}.

\begin{figure}[h!]
\begin{center}
 \begin{tabular}{c}
\includegraphics[width=9cm]{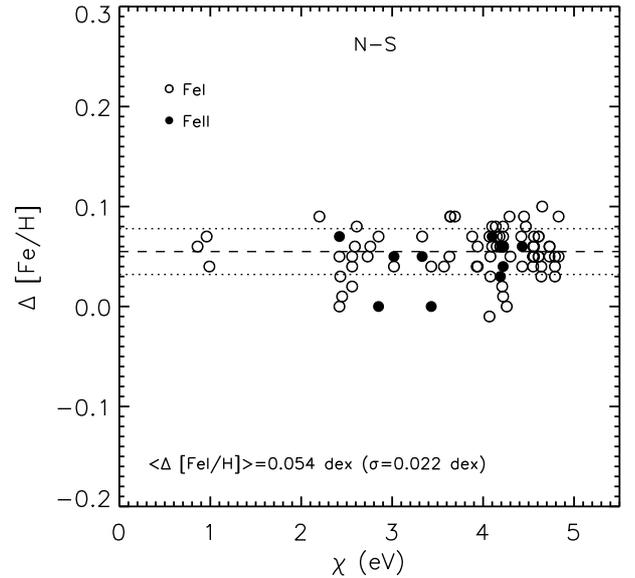}
\vspace{-.1cm}
 \end{tabular}
\caption{Iron abundance difference between XO-2N and XO-2S versus $\chi$. Dashed line shows the mean $\Delta {\rm [\ion{Fe}{i}/H]}$, while dotted lines represent the standard deviation $\pm 1 \sigma$ from the mean, where $\sigma$=0.022 dex.}
\label{fig:XO-2_diff_EP}
 \end{center}
\end{figure}

{\it iii. Differential analysis.} Here, as stellar parameters of XO-2N, we considered the values derived in the line-by-line analysis, while XO-2S was analysed differentially with respect to the companion, strictly using the same line set and EWs as in the previous step. We applied the differential abundance method widely described in \cite{grattonetal2001} and \cite{desidera04,desideraetal2006}, which yields very accurate results when, in particular, the two components are very similar in stellar parameters (as for the XO-2 components). 
In summary, the most important ingredient in this analysis is the difference in temperature between the components. In our case, we take advantage of the knowledge of the luminosity difference between the targets, as we can assume they are at the same distance from us. We considered as magnitude differences in different bands the mean values reported in Sect.\ref{sec:lcanalysis1} and as bolometric corrections those derived by using the code provided by \cite{casagrande14}, which allows to estimate the bolometric corrections in several photometric bands using T$_{\rm eff}$, [Fe/H], and log\textit{g} as primary inputs. Then, since the stars are on the main sequence, we could obtain accurate estimates of the difference $\Delta \log g$ between their surface gravities (see Eq. 6 in \citealt{desideraetal2006}, where as stellar mass difference we assumed 0.01 $M_\odot$, as given in Table \ref{table:stellarparam}). This allowed us to accurately derive their temperature difference ($\Delta T_{\rm eff}$) using the equilibrium of the iron ionization, because the difference between the abundances provided by the \ion{Fe}{i} and \ion{Fe}{ii} lines is strongly sensitive to T$_{\rm eff}$ (about 0.001 dex/K; see \citealt{grattonetal2001}). The final atmospheric parameters of the secondary were derived as differences with respect to those of the primary by means of an iterative procedure (because $\log g$, $\xi$, and [Fe/H] depend on the assumed temperatures). We interrupted the procedure considering reached the convergence when the grid step-size was of 1 K in T$_{\rm eff}$, 0.001 dex in $\log g$, and 0.01 km/s in $\xi$. No other $\sigma$ clipping was implemented here. The final adopted differences in stellar parameters are listed in Table~\ref{tab:atmospheric_parameters}, while Fig.~\ref{fig:XO-2_diff_EP} shows $\Delta${\rm [Fe/H]} as a function of the excitation potential. The under-abundance in iron and the higher effective temperature of XO-2S when compared to XO-2N are 
confirmed by this procedure. All differential parameters we obtained are affected by both internal and systematic errors. Internal error on $\Delta T_{\rm eff}$ was derived from the line-by-line scatter of \ion{Fe}{i} and \ion{Fe}{ii} and the errors in the other atmospheric parameters. Internal error in $\Delta \log g$ includes the uncertainties in the differences in mass, magnitude, bolometric correction, and effective temperature. Error in the microturbulence difference was obtained by summing in quadrature the error contributions in $\xi$ of each component. The error in the iron abundance difference was derived taking into account the uncertainties in all stellar parameters added quadratically. All errors in the differential analysis are very low, as summarized in Table~\ref{tab:atmospheric_parameters}, and this demonstrates how this procedure is efficient to unveil small differences in atmospheric parameters, removing many sources of systematic errors. In fact, as remarked by \cite{desideraetal2006}, external uncertainties, due to, e.g., wrong estimations of parallax and mass, or inadequacies of model atmospheres and LTE deviations, have negligible effects on our results, because of the similar characteristics of the components. 
In the end, the difference in iron abundance of XO-2N with respect to XO-2S is of $+0.054$~dex at a more than $3\sigma$ level.

The difference in iron abundance between the two XO-2 stellar components poses an interesting issue. They belong to a visual binary and, as normally assumed for such systems, they should share the same origin and initial bulk metallicity. A relevant characteristic of this system is that both of the stars host planets. Thus, for components of wide binaries where at least one star has a planet, a reasonable hypothesis to explain any measured and significant difference in their present-day elemental abundances is that most likely the planet formation process had played a relevant role. The higher iron abundance of XO-2N when compared to XO-2S might be due to past ingestion of dust-rich or rocky material, coming from the inner part of the proto-planetary disk and pushed into the host star by the hot Jupiter XO-2N as it migrated inward to its current orbit. A second mechanism, which acts on a different time scale and after the pre-main-sequence stage, is presented by \cite{fabrycky07}, who discuss the pollution of the stellar photosphere with metals produced by mass loss of inward migrating hot Jupiters. Also, with their simulations \cite{kaib13} showed that very distant binary companions may severely affect planetary evolution and influence the orbits of any planet around the other component of the system, thus maybe favouring the ingestion of material by the host star. Following the results shown in Fig. 2 of \cite{pinsonneault01}, we roughly estimated that XO-2N could have ingested an amount of iron slightly higher than 5 $\mearth$ to increase its photospheric iron content by $\sim$0.05 dex, given its T$_{\rm eff}\sim$5300 K.
The difference in the iron abundance of the two XO-2 stellar components is similar to what found by \cite{ramirez11} for the solar twins 16~Cyg~A and 16~Cyg~B ($\Delta$[Fe/H]=$0.042\pm0.016$ dex). These stars have masses close to those of the XO-2 companions and, together with a third companion 16~Cyg~C, are members of a hierarchical triple system, where the A and C components form a close binary with a projected separation $\sim$70 AU. The component B is known to host a giant planet in a long-period and highly eccentric orbit (P$\sim$800 days and e$\sim$0.69)\footnote{http://exoplanet.eu}, but unlike the case of XO-2 it is the star with a lower iron content. To explain this deficiency, \cite{ramirez11} suggest that an early depletion of metals happened during the formation phase of the 16 Cyg Bb planet. An alternative and suggestive possibility is that 16~Cyg~A may have ingested a massive planet that enriched the star with iron, while the large orbit of 16 Cyg Bb (with semi-major axis of 1.68 AU) likely prevented any mass loss from the planet. We note that, despite $\Delta$[Fe/H] is similar with that we found for XO-2, the amount of iron involved should be different. In fact, the XO-2 stars are cooler than 16~Cygni~A and B and have therefore more massive convective envelopes, implying that more iron should be necessary to pollute the XO-2N photosphere and produce almost the same $\Delta$[Fe/H]. Also \cite{lawsgonzalez2001} found the primary of the 16-Cyg system enhanced in Fe relative to the secondary. Similar studies conducted in binary stars hosting planets (see, e.g., \citealt{grattonetal2001,desidera04,desideraetal2006,schuleretal2011,teske13,liuetal2014,mack14}) did not find relevant differences in elemental abundance among the components. All these results imply that the presence of giant planets does not necessarily imply differences in the chemical composition of the host star. 

Our results will produce a more clear picture when several iron-peak, $\alpha$-, s-process, and other odd/even-Z elemental abundances will be analysed for both stars and their behaviour with the condensation temperature studied with the aim to investigate possible selective accretion of planetary material. One of the advantages of comparing coeval stars in wide binaries is that any difference in their abundance trend could be more related to accretion of rocky-planetary material rather than to the Galactic chemical evolution. This analysis will be the subject of a forthcoming paper of the GAPS series (Biazzo et al., in prep.), where we will investigate if the presence of hot Jupiters could lead the host stars to ingest material, which in turn may leave measurable chemical imprints in their atmospheric abundances.
	
	\subsection{XO-2 distance and galactic space velocities}
	\label{sec:stellarpar3} 
Using our stellar parameters, we derived an estimate of the spectroscopic distances of XO-2N and XO-2S by means of the following procedure. 
We generated Monte-Carlo (MC) normal distributions for each spectroscopic parameter T$_{\rm eff}$, [Fe/H], and log\textit{g}, composed of 10,000 random values and centred on the best estimates (Table \ref{table:stellarparam}). By keeping the stellar radii fixed to the values listed in Table \ref{table:stellarparam} (for XO-2N we used the most accurate estimate), for each MC simulation we first determined the stellar bolometric luminosity \textit{L$_{\rm *}$} (in solar units) from the Stefan-Boltzmann law, and then we derived the absolute bolometric magnitude \textit{M$_{\rm bol}$} from the relation M$_{\rm bol}$ = 4.75 -2.5$\cdot\log$(L$_{\rm *}$). By estimating the appropriate bolometric correction (BC), a value for the absolute magnitude in \textit{V}-band \textit{M$_{\rm V}$} was then obtained. The BC term was evaluated using the code provided by \cite{casagrande14}. An additional input is the color excess \textit{$E(B-V)$} of the star, which we derived through the relation $E(B-V)$ = A$_{\rm V}$(s)/3.1, where \textit{A$_{\rm V}$}(s) is the interstellar dust extinction in V-band integrated at the distance \textit{s} of the star (in pc) and measured along the line-of-sight. We derived A$_{\rm V}$(s) by adopting a simplified model of the local distribution of the interstellar dust density \citep{drimmel01}, expressed by the relation \textit{$\rho$}=\textit{$\rho_{\rm 0}$}$\cdot$sech$^{2}$(\textit{z/h$_{\rm s}$}), where \textit{z} is the height of the star above the Galactic plane and \textit{h$_{\rm s}$} is the scale-height of the dust, for which we adopted the value of 190 pc. The term \textit{z} is related to the distance \textit{s} and the Galactic latitude of the star \textit{b} by the formula \textit{z}=\textit{s}$\cdot\sin$\textit{b}. From this model we obtained the relation \textit{A$_{\rm V}$}(s) = \textit{A$_{\rm V}$}(tot)$\cdot$sinh(\textit{s}$\cdot\sin$\textit{b}/\textit{h$_{\rm s}$})/cosh(\textit{s}$\cdot\sin$\textit{b}/\textit{h$_{\rm s}$}), where 
\textit{A$_{\rm V}$}(tot) is the interstellar extinction in V-band along the line-of-sight integrated through the Galaxy, and can be estimated from 2-D Galactic maps. For this purpose we used the value \textit{A$_{\rm V}$}(tot)=0.16 mag derived from the maps of \cite{schlafly11}\footnote{available at http://irsa.ipac.caltech.edu/applications/DUST/}. By assuming s=150 pc as a prior distance of the stars \citep{burke07}, we obtained \textit{A$_{\rm V}$}(150pc)$\sim$0.06 mag, corresponding to $E(B-V)$=0.019 mag. This is the value used as input to the code of \cite{casagrande14} to obtain a first guess of the BC in \textit{V}-band. This in turn was used in the distance modulus formula \textit{V}-(M$_{\rm bol}$-BC$_{\rm V}$)=5$\cdot\log$(s)-5-A$_{\rm V}$(s), to obtain a new value for the stellar distance \textit{s}. The new distance was used to repeat the procedure iteratively, by determining at each step a new value of \textit{A$_{\rm V}$}(s) and \textit{BC$_{\rm V}$}(s), and finally another estimate of \textit{s}. When the absolute difference between the last and previous calculated values of \textit{s} was below 0.1 pc, the iterative process was interrupted and the last derived value for \textit{s} was assumed as the distance of the star for the N-\textit{th} Monte-Carlo simulation. 
The adopted estimates for the distance of the XO-2 components are the median of the distributions of the 10,000 MC values, and the asymmetric error bars defined as the 15.85$^{th}$ and 68.3$^{th}$ percentile (see Table \ref{table:stellarinfo}). Being model-dependent, we do not argue here whether the difference of $\sim$2.5 pc between the XO-2S and XO-2S distances is real. We only note that the two values are compatible within the uncertainties and that our best estimate for the distance of XO-2N locates the star few parsec closer than reported by \cite{burke07}. 

Distances, together with the stellar equatorial coordinates, the proper motions, and the average radial velocities (Table \ref{table:stellarinfo} and \ref{table:radvelxo2n}), were used to provide new estimates of the galactic space velocities (U,V,W) of the stars with respect to the Local Standard of Rest (LSR). We thus assumed (i) a left-handed system of reference (i.e. the velocity component U is positive toward the direction of the Galactic anti-center), (ii) the transformation matrix from equatorial to galactic coordinates taken from the Hipparcos catalogue \citep{perryman97}, and (iii) the correction for the solar motion in the LSR derived from \cite{coskunoglu11}. The results are listed in Table \ref{table:stellarinfo}. 
\cite{burke07} discussed the kinematics of the XO-2 system by using data for XO-2N only, and concluded that, despite the high galactic space velocity, which indicates a thick disk membership, the super-solar metallicity is indeed representative of a thin disk membership. This suggests that the high-proper motion of the system can be related to an eccentric orbit with a maximum height above the galactic plane of $\sim$100 pc, then confined to the galactic disk. 

We performed a new analysis of the galactic orbits, based on the work of \cite{barbieri02}, and we confirm the previous findings. Results for a sample of 1000 orbits show that the XO-2 system belongs to the thin disk, with a likely maximum height above the galactic plane of 120 pc. This result does not change significantly unless the distance of the stars is at least 20 pc more or less our estimates. Our analysis also indicates a possible eccentric orbit, with a modal eccentricity equal to 0.42$^{+0.09}_{-0.06}$ (the asymmetric errorbars are defined as the 10$^{th}$ and 90$^{th}$ percentile), and a minimum distance from the galactic center of \textit{R$_{\rm min}$}=3.5$\pm$1.5 kpc. If we assume \textit{R$_{\rm min}$} as a rough estimate of the galactic region where the two stars formed, this could explain their super-solar metallicity, because regions close to the galactic center are more metal rich than the solar neighbourhood. Moreover, this could also have modified the orbit of the binary with time due to the influence of the more dense stellar environment close to the galactic center, with important consequences on the evolution of the orbits of the planets.

	\begin{table}
	\begin{small}
    \caption{Basic information about the XO-2 stellar binary system}
    \label{table:stellarinfo}
	\centering
	\begin{tabular}{cccc}
	\hline
    \noalign{\smallskip}
    Parameter			 & XO-2N & XO-2S & Ref. \\
    \noalign{\smallskip}
    \hline
    \noalign{\smallskip}
	 
	UCAC4 ID	&  702-047113  & 702-047114 & (1) \\ [5pt]
   	R.A. (J2000) & 07:48:06.471  & 07:48:07.479 & (1) \\ [5pt]
    DEC (J2000)  & +50:13:32.91  & +50:13:03.25 & (1) \\ [5pt]
    Near UV (mag) & 18.153$\pm$0.047 &	17.988$\pm$0.044 & (2) \\ [5pt]    
    B (mag) & 12.002$\pm$0.015  & 11.927$\pm$0.015  & (3) \\ [5pt]
    V (mag) & 11.138$\pm$0.026 &	11.086$\pm$0.025 & (3) \\ [5pt]
    R$_{c}$ (mag) & 10.669$\pm$0.020 & 10.631$\pm$0.019 & (3) \\ [5pt]
    I$_{c}$ (mag) & 10.243$\pm$0.012 & 10.216$\pm$0.013 & (3) \\ [5pt]
    J (mag)	&  9.744$\pm$0.022 &	 9.742$\pm$0.022 & (4) \\ [5pt]
    H (mag)	&  9.340$\pm$0.026 &  9.371$\pm$0.027 &	(4) \\ [5pt]
    K$_{s}$ (mag) & 9.308$\pm$0.021 & 9.272$\pm$0.021 & (4) \\ [5pt] 
    Distance (pc) & 145.5$^{+2.9}_{-3.0}$ & 148.0$^{+2.8}_{-2.9}$ & (5) \\ [5pt]
    $\mu_{\alpha}$ (mas yr$^{-1}$) & $-$31.9$\pm$3.0 & $-$29.9$\pm$3.6 & (1) \\ [5pt] 
     $\mu_{\delta}$ (mas yr$^{-1}$) & $-$156.2$\pm$3.3 & $-$156.2$\pm$3.6 & (1) \\ [5pt] 
     U (km s$^{-1}$)  & 70.39$^{+0.79}_{-0.81}$  &  71.05$^{+0.75}_{-0.76}$ & (5) \\ [5pt]
     V (km s$^{-1}$)  & $-$76.0$^{+2.0}_{-2.0}$  &  $-$77.65$^{+1.92}_{-1.88}$ & (5) \\ [5pt]
     W (km s$^{-1}$)  & $-$2.91$^{+0.67}_{-0.66}$ &  $-$3.46$^{+0.64}_{-0.62}$ & (5) \\ 
     \noalign{\smallskip}
     \hline 
   \noalign{\smallskip}
     & \multicolumn{2}{c}{Relative magnitudes (XO-2N - XO-2S)} & \\ 
     \hline 
\noalign{\smallskip}
    $\Delta$B (mag) & \multicolumn{2}{c}{0.075$\pm$0.002} & (6) \\ [5pt] 	
    $\Delta$V (mag) & \multicolumn{2}{c}{0.057$\pm$0.002} & (6) \\ [5pt] 
    $\Delta$R$_{c}$ (mag) & \multicolumn{2}{c}{0.040$\pm$0.002}  & (6) \\ [5pt] 
                          & \multicolumn{2}{c}{0.041$\pm$0.004}  & (7) \\ [5pt] 
    $\Delta$r$^{\prime}$ (mag)    &  \multicolumn{2}{c}{0.038$\pm$0.001}  & (8) \\ [5pt]                    
    $\Delta$I$_{c}$ (mag) & \multicolumn{2}{c}{0.031$\pm$0.003} & (6) \\ [5pt]
                          & \multicolumn{2}{c}{0.030$\pm$0.006} & (9) \\ 
    \noalign{\smallskip}                       
    \hline
	\end{tabular}
	\tablebib{(1) UCAC4 catalogue \citep{zacha13}; (2) \cite{bianchi11}; (3) APASS all-sky survey \cite{henden14} ; (4) 2MASS catalogue \citep{skrut06}; (5) This work; (6) This work (Serra La Nave Observatory); (7) This work (TASTE project); (8) \citep{kundurthy13}, $\lambda_{\rm 0}$=626 nm, 4 epochs ; (9) This work (APACHE project).
    }
	\end{small}
	\end{table}


    \begin{table}
  \begin{small}
    \caption{Stellar parameters for the XO-2 components derived from the analysis of the HARPS-N spectra and from stellar evolutionary tracks. The adopted values represent the weighted mean of individual measurements obtained with different methods (except for the projected velocity \textit{V}sin\textit{I$_{\rm \star}$}, for which we adopt the measurement derived from the Rossiter-McLaughlin effect), and their associated uncertainties are, for a more conservative estimate, the average of the uncertainties of the individual values.}
    \label{table:stellarparam}
	\centering
	\begin{tabular}{cccc}
	 \hline
         \noalign{\smallskip}
        Parameter 	  & XO-2N & XO-2S & Note \\  
         \noalign{\smallskip}
         \hline
         \noalign{\smallskip}
    T$_{eff}$ [K] & 5332$\pm$57 & 5395$\pm$54 & \\ [3pt]

	$\log g$ [cgs] & 4.44$\pm$0.08  & 4.43$\pm$0.08 & \\ [3pt]

	[Fe/H] [dex]  & 0.43$\pm$0.05 & 0.39$\pm$0.05 & \\ [3pt]

    Microturb. $\xi$ [km s$^{-1}$]  & 0.88$\pm$0.11  & 0.90$\pm$0.10 & \\ [3pt]

	\textit{V}sin\textit{I$_{\rm \star}$} [km s$^{-1}$] & 1.07$\pm$0.09 & 1.5$\pm$0.3 & \\ [3pt]

	Mass [M$_\odot$]     & 0.97$\pm$0.05 & 0.98$\pm$0.05 & (1)  \\ [1pt]
					     & 0.96$\pm$0.05 & - & (2) \\ [3pt]
	Radius [R$_\odot$]   & 1.01$^{+0.1}_{-0.07}$  & 1.02$^{+0.09}_{-0.06}$  & (1)\\ [3pt]
					     & 0.998$^{+0.033}_{-0.032}$ & - & (2) \\ [3pt]
	Age [Gyr]   	         & 7.9$^{+2.3}_{-3.0}$ & 7.1$^{+2.5}_{-2.9}$ & (1)\\ [3pt]
				         & 7.8$^{+1.2}_{-1.3}$ & - & (2) \\ [3pt]
	Luminosity [L$_\odot$] &	 0.70$\pm$0.04	& 0.79$\pm$0.14 & for XO-2N: (2) \\ [3pt] 
	                       &    &   &  for XO-2S: (1)  \\  
	  \noalign{\smallskip}
      \hline
	\end{tabular}
	\tablefoot{
	 (1) Matching the T$_{\rm eff}$, [Fe/H], and $\log g$ to the Yonsei-Yale evolutionary tracks. 
	 (2) Matching the T$_{\rm eff}$, [Fe/H], and stellar density to the Yonsei-Yale evolutionary tracks.
	 } 
    \end{small}   
    \end{table}
   
\begin{table}
\caption{Stellar atmospheric parameters and their differences for XO-2N and XO-2S. ee discussion in Sect.\ref{sec:stellarpar2}} 
\label{tab:atmospheric_parameters}
\centering
\begin{tabular}{cccc}
\hline
\noalign{\smallskip}
Parameter     &   XO-2N & XO-2S & $\Delta_{\rm N-S}$ \\
\noalign{\smallskip}
\hline
\noalign{\smallskip}
\multicolumn{4}{c}{\it Analysis based on equivalent widths } \\
T$_{\rm eff}$ [K]    & 5320$\pm$50     & 5330$\pm$50    & $-10\pm71$  \\ 
$\log g$ [cgs]       & 4.46$\pm$0.08   & 4.41$\pm$0.12  & $+0.05\pm0.14$\\  
$\xi$ [km s$^{-1}$]         & 0.89$\pm$0.14   & 0.93$\pm$0.04  & $-0.04\pm0.15$\\ 
$[$Fe/H$]^a$ [dex]   & 0.37$\pm$0.07   & 0.32$\pm$0.08  & $+0.05\pm0.11$\\ 
\noalign{\smallskip}
\multicolumn{4}{c}{\it Line-by-line analysis}\\
T$_{\rm eff}$ [K]    & 5290$\pm$18     &  5322$\pm$25   & $-32\pm31$  \\ 
$\log g$ [cgs]       & 4.43$\pm$0.10   & 4.41$\pm$0.10  & $+0.02\pm0.14$\\  
$\xi$ [km s$^{-1}$]         & 0.86$\pm$0.06   & 0.93$\pm$0.05  & $-0.07\pm0.08$\\ 
$[$Fe/H$]^a$ [dex]   & 0.37$\pm$0.07   & 0.32$\pm$0.08  & $+0.05\pm0.11$\\ 
\noalign{\smallskip}
\multicolumn{4}{c}{\it Differential analysis}\\
T$_{\rm eff}$ (K]    & 5290$\pm$18     & 5325$\pm$37     & $-35\pm8$  \\ 
$\log g$ [cgs]       & 4.43$\pm$0.10   & 4.420$\pm$0.094 & $+0.010\pm0.020$\\  
$\xi$ [km s$^{-1}$]        & 0.86$\pm$0.06   & 0.93$\pm$0.03   & $-0.07\pm0.07$\\ 
$[$Fe/H$]$\tablefootmark{a} [dex]   & 0.37$\pm$0.07   & 0.32$\pm$0.08   & $+0.054\pm0.013$\\[3pt] 
\hline
\end{tabular}
\tablefoot{
\tablefoottext{a}{The iron abundance [Fe/H] refers to the abundance of \ion{Fe}{i}.}}
\end{table}

\section{Analysis of the light curves}
\label{sec:lcanalysis}
	\subsection{Out-of-transit photometry}
	\label{sec:lcanalysis1}
\begin{figure*}
   \centering
   \includegraphics[width=15cm]{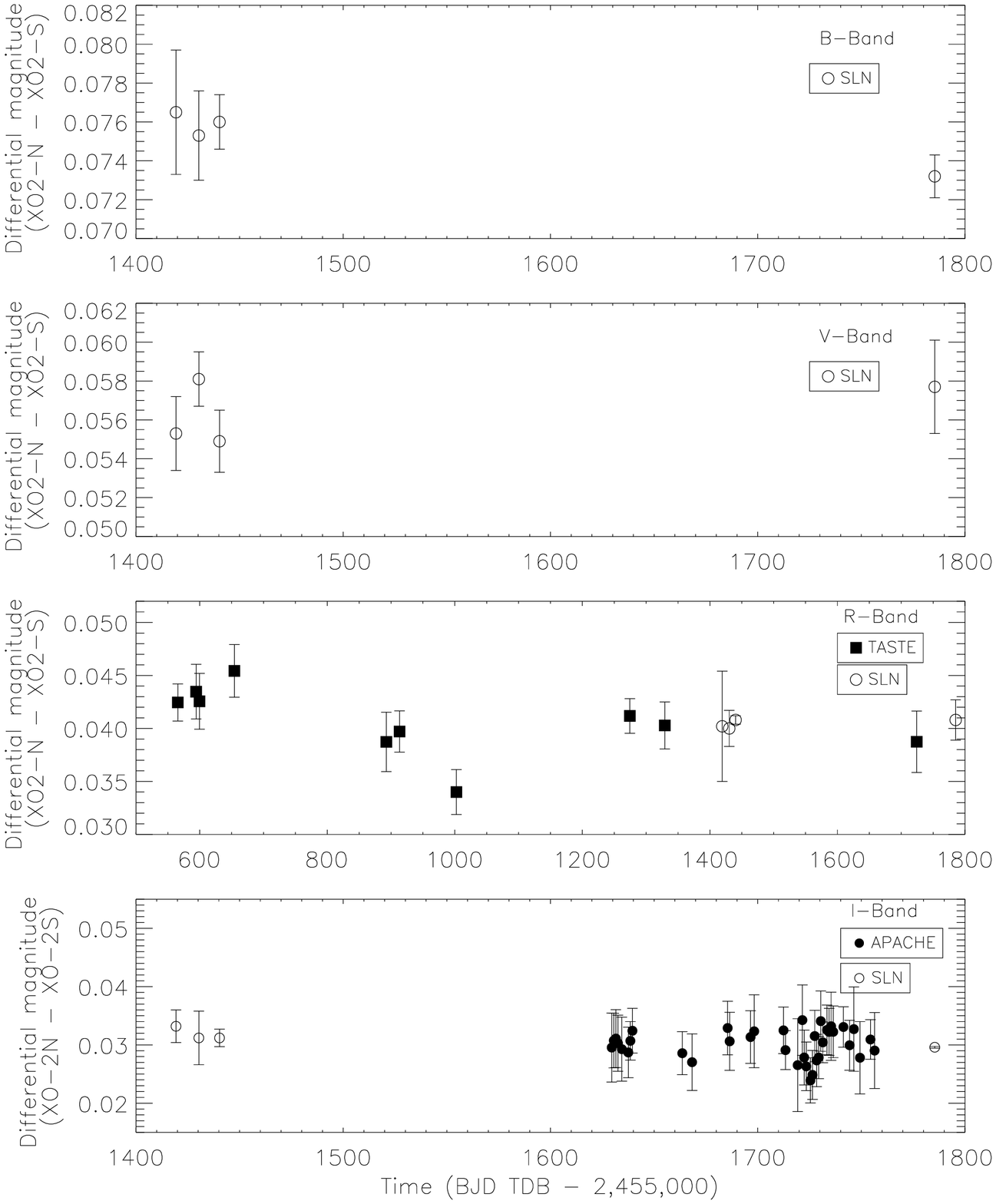}
   \caption{Magnitude differences between the star XO-2N and its companion XO-2S as measured in \textit{BVRI} passbands from data of the TASTE project, APACHE survey, and Serra La Nave Observatory. These differences were evaluated from out-of-transit photometry.}
   \label{fig:relativephotometry}%
    \end{figure*}

Figure \ref{fig:relativephotometry} shows the relative photometry of the two components of the XO-2 system obtained from different facilities. Measurements were collected in the \textit{BVRI} standard photometric passbands and the points corresponding to the transits of the planet XO-2Nb were not included in the calculation of the intra-night average values. The uncertainties associated to the measurements are the RMS of the points collected in each single night of observation, and when only one measurement per night was available from SLN, we adopted the dispersion of all the data as the error. The average values for the magnitude differences in each band are indicated in Table \ref{table:stellarinfo}. We note that the independent measurements in \textit{R} (TASTE and SLN) and \textit{I} (APACHE and SLN) bands agree well each other. Our derived relative magnitudes have been used in the differential abundance analysis discussed in Sect.\ref{sec:stellarpar2}.

\begin{figure}
   \centering
   \includegraphics[width=9.5cm]{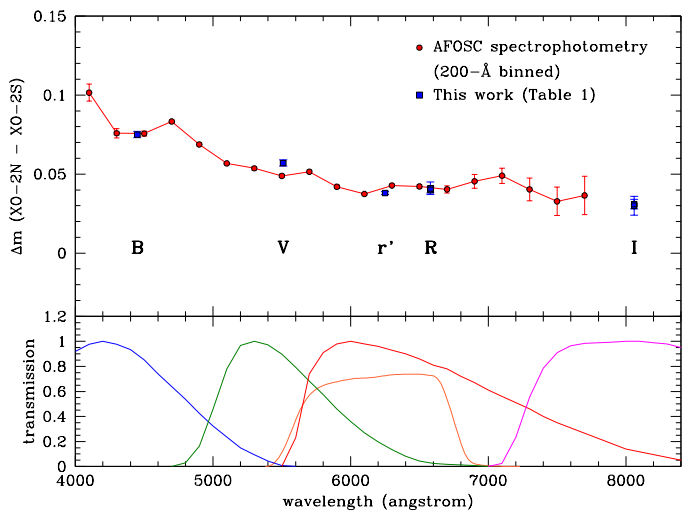}
   \caption{\textit{Upper panel}. Differential spectro-photometry of the XO-2 components obtained with the AFOSC camera at the 182cm Asiago telescope. Red dots represent the mean values of measurements in bin of 200 $\AA$, while blue squares indicate the broad-band differential magnitudes presented in Table \ref{table:stellarinfo}. \textit{Lower panel}. Bandpasses of the filters used for the broad-band measurements.}
   \label{fig:afoscdiffspectrum}%
    \end{figure}

On the night between October 31, and November 1 2014, we took for 30 minutes a series of 60 s exposures of the XO-2 field with the Asiago Faint Object Spectrograph and Camera (AFOSC) at the 182cm Asiago Astrophysical Observatory telescope, while the field was close to the zenith. By placing both components in a 8$^{\prime\prime}$ slit and using a grism covering the spectral range 3900-8000 $\AA$, we obtained a final simultaneous differential spectrum by dividing each other the combined individual spectra in each wavelength resolution element. The differential spectrum is shown in Fig. \ref{fig:afoscdiffspectrum}. These observations were characterised by a very good instrumental stability and turned out to be useful to strengthen the results obtained with broad-band measurements of the differential magnitudes. In fact, these are in good agreement with the AFOSC measurements, as shown in Fig. \ref{fig:afoscdiffspectrum}. Due to the high scatter of the data in the \textit{I}-band caused by fringing, we did not include in the plot the measurements with wavelengths greater than $\sim$770 nm.
 
    \subsubsection{Rotational periods}
	\label{sec:rotperiods} 
 
\begin{figure}
   \centering
   \includegraphics[width=9.5 cm]{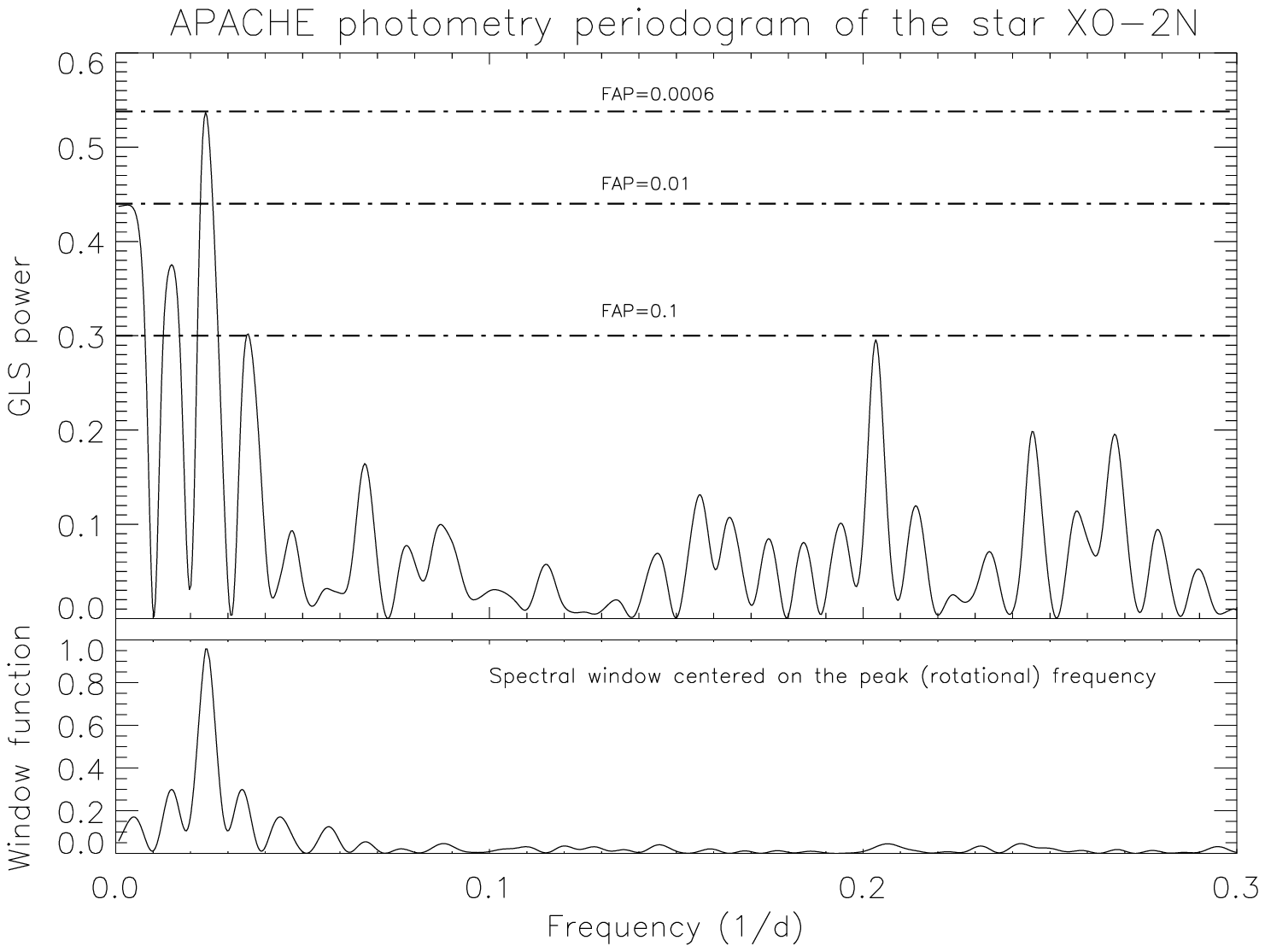}
   \includegraphics[width=8.7cm]{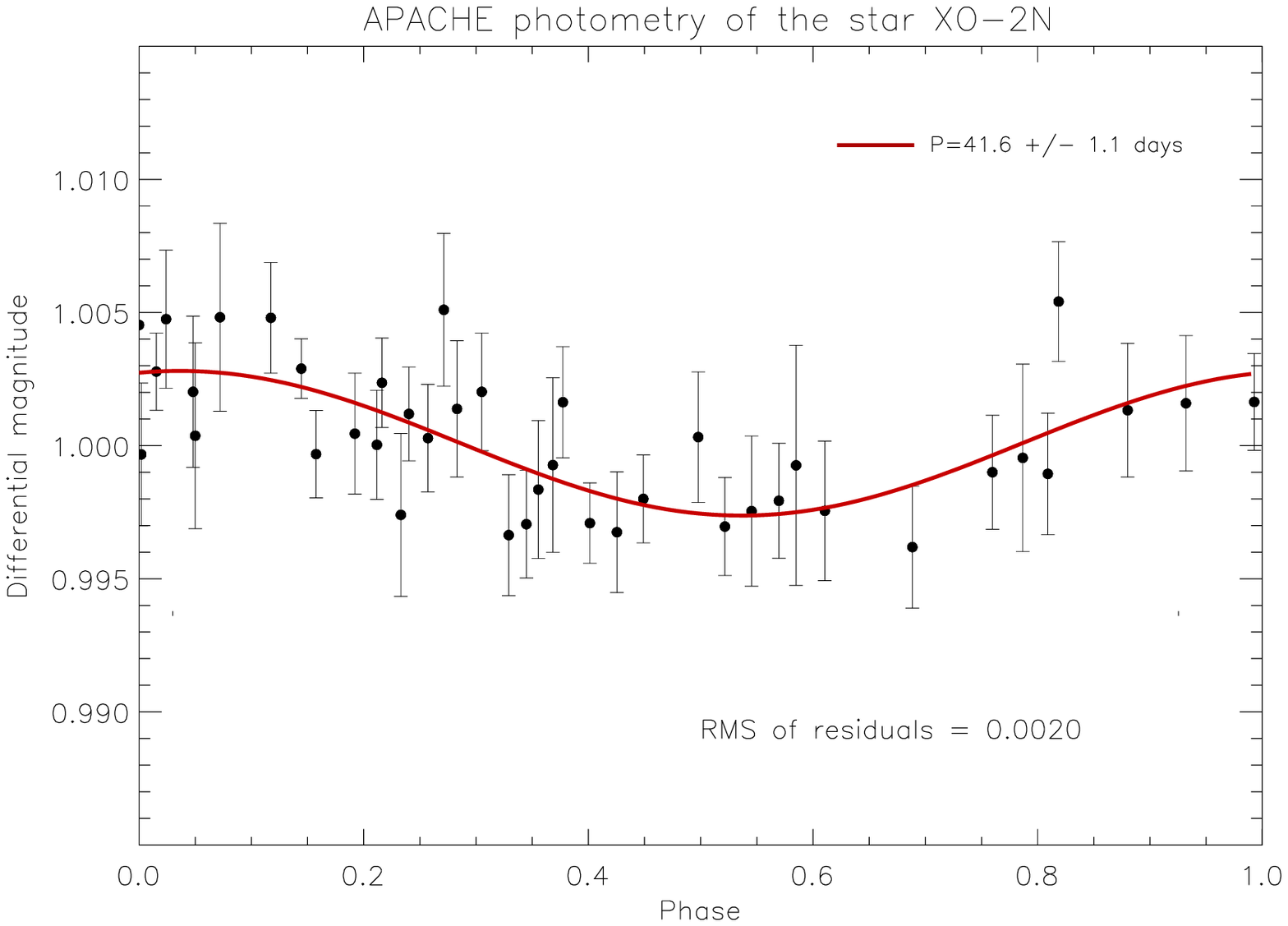}
   \caption{\textit{Upper panel}. The GLS periodogram of the XO-2N light curve collected by the APACHE survey 
   and the corresponding window function, with the highest peak translated to the frequency of the best significant peak found in the periodogram. Levels of different FAP are also indicated. \textit{Lower panel}. The APACHE light curve of XO-2N folded at the best period found by GLS with superposed the best sinusoidal fit of semi-amplitude A=0.0027 mag (solid red line).}
   \label{fig:apachefolded1}
    \end{figure}

\begin{figure}
   \centering
   \includegraphics[width=9.5cm]{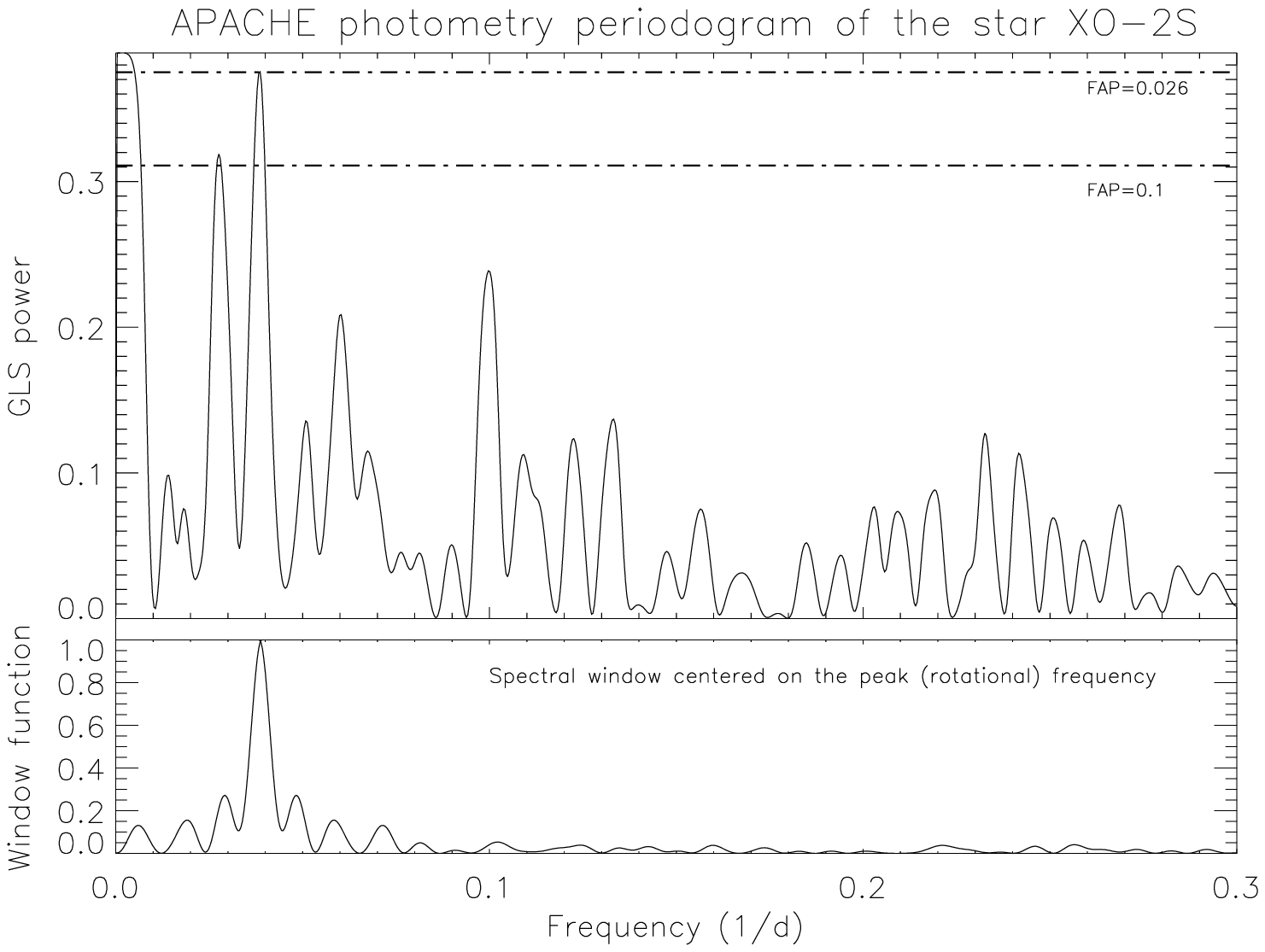}
   \includegraphics[width=8.7cm]{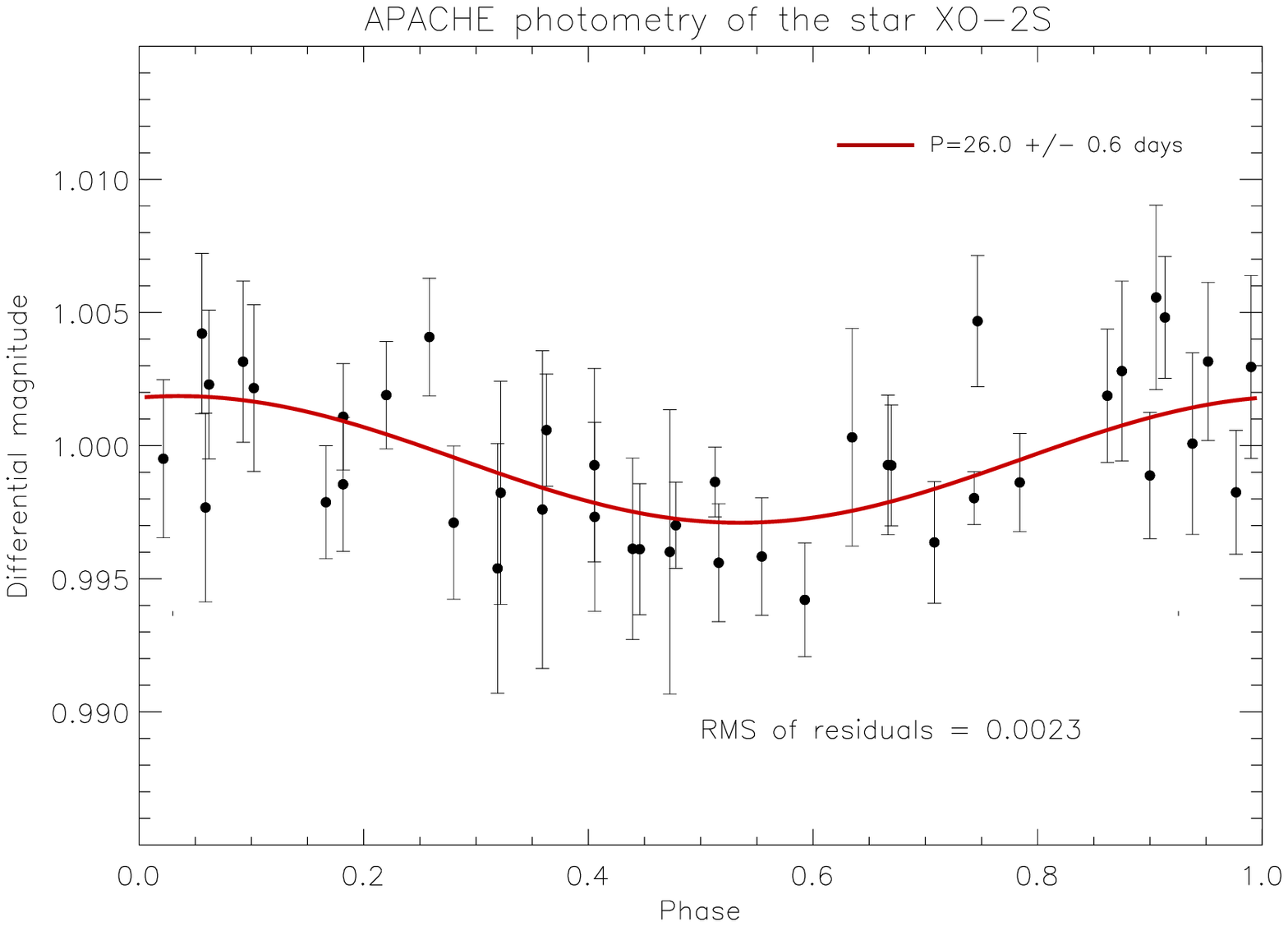}
   \caption{\textit{Upper panel}. The GLS periodogram of the XO-2S light curve collected by the APACHE survey and the corresponding window function, with the highest peak translated to the frequency of the best significant peak found in the periodogram. Levels of different FAP are also indicated. \textit{Lower panel}. The APACHE light curve of XO-2S folded at the 
   best period found by GLS with superposed the best sinusoidal fit of semi-amplitude A=0.0024 mag (solid red line).}
   \label{fig:apachefolded2}%
    \end{figure}

Thanks to the timespan of the APACHE \textit{I}-band observations and $\sim$800 useful data points for each component, we could analyse the photometric time series to search for periodic modulations ascribable to the rotation of the stars and produced by brightness inhomogeneities distributed on the stellar photospheres (which are assumed to change on time scales longer than the timespan of the observations). Taking into account that the XO-2 system is quite old (see Table \ref{table:stellarparam}) and the two stars are magnetically quiet (see discussion in Sect.\ref{sec:stellaractivity}), they are expected to have rotation periods of some tens of days and low amplitudes in the light curve variations. We used the Generalized Lomb-Scargle algorithm (GLS; \citealt{zechmeister09}) to search for periodicities in our photometric time series, after binning in a single point the data collected during each night. The data are available on-line at the CDS, each containing the timestamps, differential magnitudes and corresponding uncertainties. Results are shown in Fig. \ref{fig:apachefolded1} and \ref{fig:apachefolded2} for XO-2N and XO-2S. The major peaks at low frequencies in the periodograms are due to the limited time baseline of our observations.
For the North component we find a peak in the periodogram corresponding to a period P=41.6$\pm$1.1 days, with a semi-amplitude of the folded light curve of 0.0027$\pm$0.0004 mag. Because there is no evidence of harmonics in the periodogram, the measurements can be arranged almost perfectly with a sine function. To estimate the significance of the best period, we performed a bootstrap analysis (with replacement) by using 10,000 fake datasets derived from the original photometry, and in six cases we found a power greater than that associated to the best period in the real dataset. This corresponds to a False Alarm Probability (FAP) equal to 0.06$\%$, which makes the result for XO-2N rather robust. 

Interestingly, to our knowledge other two systems with an old star have measured rotation periods thanks to \textit{Kepler} data. The A and B components of the 16 Cygni system, which is estimated to be 6.87$\pm$2 Gyr old, have rotation periods 23.8$^{+1.5}_{-1.8}$ and 23.2$^{+11.5}_{-3.2}$ days respectively \citep{davies15}. Another example is that of Kepler-77, a star-planet system 7$\pm$2 Gyr old. The detected stellar rotation period is 36$\pm$6 days, with sizeable changes in the shape of the light curve modulation \citep{gandolfi13}. This case and especially that of 16 Cyg B, with  uncertainties of several days, demonstrate that for old late-type stars the determination of the true rotation period can be particularly difficult. The intrinsic evolution of their surface active regions takes place on a time scale shorter than the rotation period, and therefore starspots become bad tracers of stellar rotation. In the Sun, depending on the phase of the solar cycle, the rotation period measured by fitting a sinusoid to the total solar irradiance - a good proxy for the Sun-as-a-star optical modulation - varies between 22.0 and 31.9 days (the true synodic period being 27.27 days). The variations are especially large during the maximum of the solar cycle because several large active regions are simultaneously present on the solar surface and evolve on time scales of 1-2 weeks, i.e. significantly shorter than the true rotation period (e.g., \citealt{lanza03}).
 
Even if the rotation period of XO-2N appears well constrained, however the complications expected for old late-type stars could actually have prevented us to detect an accurate value. This is indeed the case for the companion star XO-2S, for which a less convincing detection resulted in our data. The GLS periodogram returned the highest peak signal at $f$=0.0384 days$^{-1}$, corresponding to P=26.0$\pm$0.6 days, with a semi-amplitude of 0.0024$\pm$0.0005 mag of the folded light curve. Also for XO-2S there is no evidence of harmonics. The bootstrap analysis in this case led to a FAP=2.6$\%$, which casts some doubts on the reality of the signal, even if the period is still compatible with the measured projected velocity \textit{V}sin\textit{I$_{\star}$}. The similarity between the observed periodogram and the recentered spectral window (bottom panel in  the upper part of Fig. \ref{fig:apachefolded2}) strengthens our confidence on the stellar origin of the periodicity we found. However, it must be noticed a second peak at 0.029 days$^{-1}$ (P=34.5 days) with a slightly less power. This has some relevance in the discussion, since a rotational period of 34.5 days matches the gyrochronology relations (e.g. see Fig. 13 in \citealt{barnes07}) better than the one of 26 days, and it is also closer to the estimates based on measurements of the activity index \textit{R$^{\prime}_{\rm HK}$} (see Table \ref{table:actindtable}). Moreover, in analogy with the case of the Sun-as-a-star, if we assume P=34.5 days as the best detection, a true period which is a bit higher than that is nonetheless reliable. Therefore more data are necessary, possibly covering several activity cycles, to reach a more robust conclusion concerning the true rotational period of XO-2S.

	\subsection{Transit photometry of XO-2N}
	\label{sec:lcanalysis2}
Following the discovery of the hot Jupiter XO-2Nb \citep{burke07}, new transit light curves of XO-2N were collected to improve the stellar and planetary parameters and search for transit timing variations \citep{fernandez09,kundurthy13,crouzet12}, and the planet atmosphere was analysed with space-based \citep{machalek09,crouzet12} and ground-based observations \citep{sing11,sing12}. 

\begin{figure*}
   \centering
   \includegraphics[width=15cm]{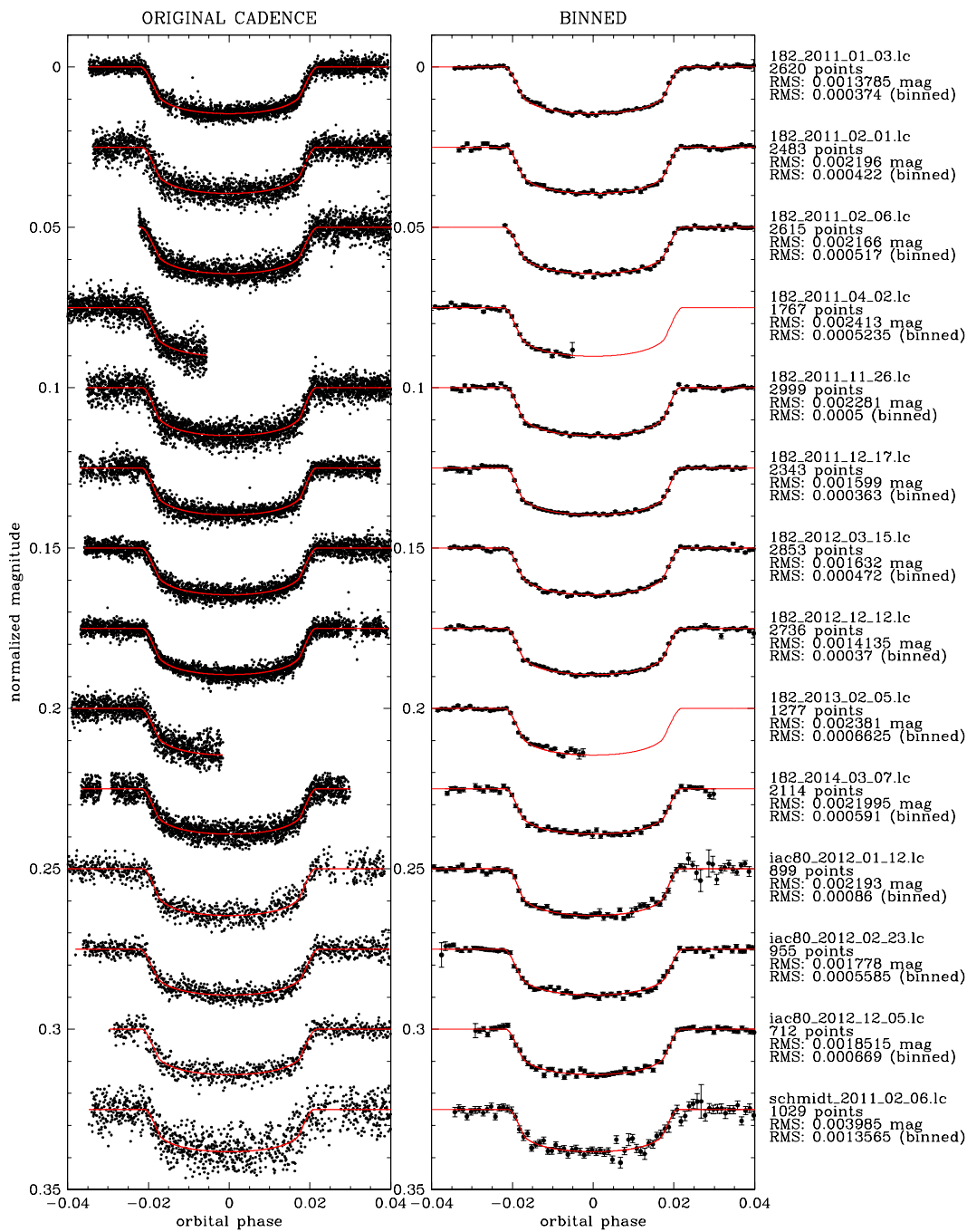}
   \caption{Transit light curves of the planet XO-2Nb collected with the Copernico (182-cm) and Schmidt 67/91-cm telescopes (Asiago Observatory) within the framework of the TASTE Project (first ten plots and the last one) and, independently, at the IAC-80 telescope (Teide Observatory, Canary Islands; remaining plots). All these observations were carried out in the Cousin/Bessell \textit{R}-band. The light curves are plotted both in their original cadence (left diagrams), and binned over 2-min intervals (right diagrams). For each transit, the RMS of the residuals from the best-fit model is also tabulated. A composite light curve has been obtained from all the individual time series and used for determining the transit parameters shown in Table \ref{table:transitparam}, as discussed in the text.}
   \label{fig:transititotale}%
    \end{figure*}

\begin{figure}
   \centering
   \includegraphics[width=9cm]{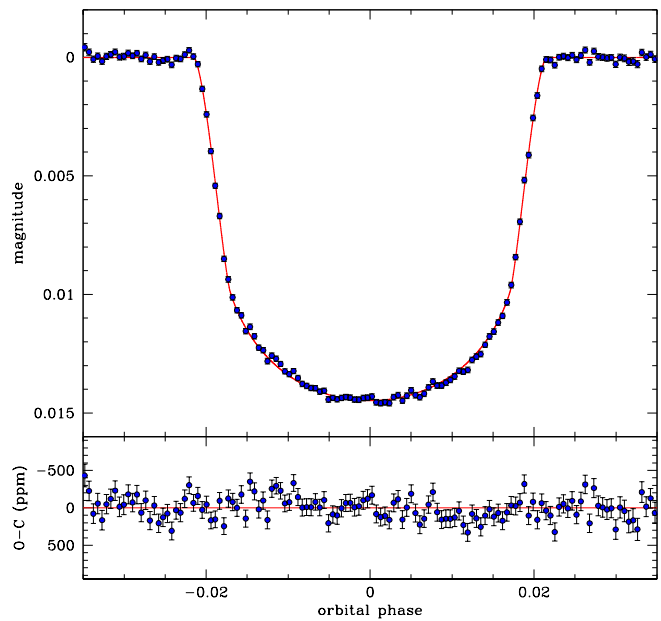}
   \caption{Photometric data presented in Fig. \ref{fig:transititotale} and combined together in a single transit light curve to appreciate the quality of the whole dataset. Data points are the weighted mean of the measurements in bins of 120 seconds, and the best-fit transit model is over plotted as a solid line, with the O-C residuals shown in the lower panel.}
   \label{fig:transitostacked}%
    \end{figure}

Through the combined analysis of 14 new transit light curves, we present here refined parameters for the XO-2N system, together with those in Table \ref{table:stellarparam}. The individual transit light curves are shown in Fig. \ref{fig:transititotale}, and the full dataset combined in a single, high-quality transit light curve is shown in Fig. \ref{fig:transitostacked}. The information about each dataset is summarized in Table \ref{table:transitdetail}.  
The raw scientific frames were first corrected for bias, dark and flat-field using standard software tools. The STARSKY pipeline (see \citealt{nascimbeni13}, and references therein) was employed to perform differential photometry between XO-2N and a set of optimally-weighted reference stars. The whole set of fourteen light curves was then simultaneously analysed with the \texttt{JKTEBOP}\footnote{http://www.astro.keele.ac.uk/jkt/codes/jktebop.html} code (version 34, \citealt{southworth04}). The adopted transit model parametrizes the stellar limb darkening (LD) with a quadratic law, i.e. with the two parameters \textit{u$_{\rm 1}$} and \textit{u$_{\rm 2}$}. Assuming a circular orbit, seven parameters were left free to vary in the fitting procedure: the period \textit{P} and reference time \textit{T$_{\rm 0}$} of the underlying transit ephemeris, the ratio \textit{k}=\textit{R$_{\rm \mathrm{p}}$/R$_{\rm \star}$} and sum (\textit{R$_{\rm \mathrm{p}}$+R$_{\rm \star}$})/\textit{a} of the fractional radii (\textit{a} being the semi-major axis of the system), the orbital inclination \textit{i}, and the LD parameters \textit{u$_{\rm 1}$} and \textit{u$_{\rm 2}$}. Leaving both \textit{u$_{\rm 1}$} and \textit{u$_{\rm 2}$} free to vary allows us to avoid to put any \emph{a priori} assumption on the stellar atmospheric parameters. Our estimates are in good agreement with those derived by interpolating the theoretical tables of \cite{claret11} using our derived stellar atmospheric parameters (0.48 and 0.21 for \textit{u$_{\rm 1}$} and \textit{u$_{\rm 2}$}). The transit model was fitted to our data with a non-linear least-squares technique; the uncertainties over the best-fit parameters are then recovered with a bootstrap algorithm, described by \cite{southworth08}. Other quantities of interest, such as the impact parameter \textit{b} or the transit duration \textit{T$_{\rm 14}$}, were derived through their analytical expressions from the fitted parameters. The final best-fit values, along with their estimated errors, are summarized in Table \ref{table:transitparam}. We give the estimates as the median of the bootstrapped distributions, with the asymmetric errorbars defined as the 15.87$^{th}$ and 84.13$^{th}$ percentile. 
Our results are in good agreement with those in the literature, in particular with the results of \cite{crouzet12} based on observations made with the\textit{ Hubble Space Telescope}. We were able to provide an updated ephemeris (with an error on \textit{P} of about 0.025 s) that we also used as a prior for the radial velocity fit (Sect.\ref{sec:radvelanalysis}). Our determination of the stellar density \textit{$\rho$/$\rho_{\rm \odot}$}, thanks to \textit{u$_{\rm 1}$} and \textit{u$_{\rm 2}$} being left free to vary, is completely independent from stellar models, and was used as input to interpolate within the Y-Y evolutionary tracks and derive an accurate estimate for the stellar mass, radius and age (Sect.\ref{sec:stellarpar}).

%

%

%

%

%

    \begin{table*}
    \caption{Log of the transit observations of XO-2Nb analysed in this work (14 distinct light curves). All of them were obtained using a Cousin/Bessell \textit{R} filter.}
    \label{table:transitdetail}
	\centering
	\begin{tabular}{lllccc}
	\hline
    \noalign{\smallskip}
    Date   &  Observatory & Telescope/instrument  &  $N$    & $\sigma$ & $\sigma_{120s}$ \\ 
                 &      &                   &         & (mmag)   & (mmag)       )\\  
    \noalign{\smallskip}
    \hline
    \noalign{\smallskip}
    
    2011/01/03 & Asiago & 182cm/AFOSC     &  2620 & 1.37 & 0.37  \\
	2011/02/01 & Asiago & 182cm/AFOSC     &  2483 & 2.19 & 0.42  \\
	2011/02/06 & Asiago & 67/91cm Schmidt &  1029 & 3.98 & 1.35  \\
	2011/02/06 & Asiago & 182cm/AFOSC     &  2615 & 2.16 & 0.51  \\
	2011/04/02 & Asiago & 182cm/AFOSC     &  1767 & 2.41 & 0.52  \\
	2011/11/26 & Asiago & 182cm/AFOSC     &  2999 & 2.28 & 0.50  \\
	2011/12/17 & Asiago & 182cm/AFOSC     &  2343 & 1.59 & 0.36  \\
	2012/01/12 & Teide  & IAC80/CAMELOT   &  899  & 2.19 & 0.86  \\
	2012/02/23 & Teide  & IAC80/CAMELOT   &  955  & 1.77 & 0.55  \\
	2012/03/15 & Asiago & 182cm/AFOSC     &  2853 & 1.63 & 0.47  \\
	2012/12/05 & Teide  & IAC80/TCP       &  712  & 1.85 & 0.66  \\
	2012/12/12 & Asiago & 182cm/AFOSC     &  2736 & 1.41 & 0.37  \\
	2013/02/05 & Asiago & 182cm/AFOSC     &  1277 & 2.38 & 0.66  \\
	2014/03/07 & Asiago & 182cm/AFOSC     &  2114 & 2.19 & 0.59  \\ \hline
	
	\end{tabular}
	\tablefoot{The columns list: the evening date, the observatory and instrument employed, the number of data points \textit{N}, the photometric RMS of the residuals from the best-fit model (both for the original cadence, \textit{$\sigma$}, and after binning over 120-s intervals, \textit{$\sigma_{120}$}).}
    \end{table*}
%


    \begin{table}
    \caption{Transit parameters derived from the combined analysis of the 14 transit light curves of XO-2Nb plotted in 
    Fig. \ref{fig:transititotale}.}
    \label{table:transitparam}
	\centering
	\begin{small}
	\begin{tabular}{ll}
	\hline
    \noalign{\smallskip}
    Parameter             &  Value \\
    \noalign{\smallskip}
    \hline
    \noalign{\smallskip}
	
	$P$ (d)               &   2.61585922$\pm0.00000028$ \\[3pt]
	$(\rplanet+\rstar)/a$ &   0.1393$\pm0.0018$ \\ [3pt]
	$\rplanet/\rstar$     &   0.10490$^{+0.00059}_{-0.00063}$ \\ [3pt]
	$(\rplanet/\rstar)^2$ &   0.01100$\pm0.00013$ \\ [3pt]
	$i$ (deg)             &   87.96$^{+0.42}_{-0.34}$ \\ [3pt]
	$a/\rstar$            &   7.928$^{+0.099}_{-0.093}$ \\ [3pt]
	$b$                   &   0.287$^{+0.043}_{-0.055}$ \\ [3pt]
	$T_{\rm 14}$ (min)        &   162.15$\pm0.36$ \\ [3pt]
	$T_{\rm 14}$ (d)          &   0.11260$\pm0.00025$ \\ [3pt]
	$T_c$ (BJD TDB-2,450,000) &   5565.546480$\pm0.000054$ \\ [3pt]
	$u_1$                 &   0.474$^{+0.030}_{-0.028}$ \\ [3pt]
	$u_2$                 &   0.171$^{+0.067}_{-0.070}$ \\ [3pt]
	$\rho/\rho_\odot$     &   0.960$^{+0.037}_{-0.034}$ \\ \noalign{\smallskip}
	 \hline
	\end{tabular}
    \tablefoot{The listed parameters are: orbital period \textit{P}, sum and ratio of the fractional radii (\textit{R$_{\rm \mathrm{p}}$+R$_{\rm \star}$})/\textit{a}, (\textit{R$_{\rm \mathrm{p}}$/R$_{\rm \star}$}), orbital inclination \textit{i}, semi-major axis \textit{a} scaled by stellar radius \textit{R$_{\star}$}, impact parameter \textit{b}, total duration of the transit from the first to the fourth contact \textit{T$_{\rm 14}$}, reference time of the ephemeris \textit{T$_{\rm 0}$} (time standard is Barycentric Julian Day calculated from UTC times: \textit{$\mathrm{BJD}_{\mathrm{UTC}}$}), the linear and quadratic LD parameters \textit{u$_{\rm 1}$}, \textit{u$_{\rm 2}$}, the average stellar density \textit{$\rho$/$\rho_{\odot}$}. 
    }
    \end{small}
    \end{table}
    

\section{Radial velocities of XO-2N}
\label{sec:radvelanalysis}

	We observed the star XO-2N as part of a program aimed at searching for additional outer planets in the system, and we investigated the Rossiter-McLaughlin (RML) effect by acquiring high-resolution spectra with HARPS-N during one transit event of the planet XO-2Nb. 


    \begin{table*}  
    \caption{Orbital parameters for the XO-2N and XO-2S systems. Errors and upper limits refer to 1$\sigma$ uncertainties. The results for XO-2Nb are discussed in Sect. \ref{sec:radveltimeseries}, while those for the XO-2S planets are taken from \cite{desidera14}. We also include our estimates of few parameters for a second companion to XO-2N, namely XO-2Nc, despite its existence is a matter of debate, as discussed in Sect. \ref{sec:radvelacceleration}.}
    \label{table:rvxoharpsn}
	\centering
	\small
	\begin{tabular}{cccccc}
	\hline
    \noalign{\smallskip}
    Parameter  & XO-2Nb & XO-2Nc\footnote{see Sect.\ref{sec:radvelacceleration} for our discussion about the existence of this companion.} & XO-2Sb & XO-2Sc & Note \\ 
    \noalign{\smallskip}
    \hline
    \noalign{\smallskip}
	
	P [days] & $2.61585922\pm2.8 \cdot 10^{-7}$ & $ \ge$ 17 years & 18.157$\pm$0.034  & 120.80$\pm$0.34 & fitted \\ [5pt]
	K	[m s$^{-1}$]   &  $90.17\pm0.82 $   &  $\ge 20$ & 20.64$\pm$0.85 & 57.68$\pm$0.69 & fitted \\ [5pt]
	$K \cdot P^{-2}  [\ms \rm day^{-2}]$ & - & $-5.13\pm0.54 \cdot 10^{-7}$  &   -    &  -  & derived \\ [5pt]
	$\sqrt{e} \sin \omega$  &  -   & - &   $-$0.314$_{-0.052}^{+0.059}$     &   $-$0.388$_{-0.012}^{+0.013}$ & fitted \\ [5pt]
	$\sqrt{e} \cos \omega$    & -     &   - &  0.282$_{-0.065}^{+0.054}$     &  $-$0.038$_{-0.034}^{+0.033}$ & fitted \\ [5pt]
	$e \cos \omega $   &   $ -0.0009_{-0.0022}^{+0.0034}$   &  - & - & - & fitted \\ [5pt]
	$e \sin \omega $   &   $ -0.0004_{-0.0061}^{+0.0040} $ &  - & - & - & fitted \\ [5pt]
	\textit{e}       &  < 0.006  &  - &  $ 0.180\pm0.035 $  &  $ 0.1528_{-0.0098}^{+0.0094}$ & derived  \\ [5pt]
	$\omega$ [deg]    &  unconstrained    &  - &  $ 311.9 \pm 9.5  $    &  $ 264.5\pm4.9 $  & derived    \\ [5pt]
	$T_c$ [$\rm BJD_{TDB}$-2,450,000]  &  $5565.546480\pm5.4 \cdot 10^{-5} $  &  - & 6413.11$_{-0.86}^{+0.82}$  &  6408.1$_{-1.9}^{+1.8}$ & fitted \\ [5pt]
	$\gamma_{\rm HDS} [\ms] $& $-17.3\pm2.0$ & - & - & - & fitted \\[5pt]
	$\gamma_{\rm HIRES} [\ms]$ & $-13.5\pm2.9$ & - & - & - & fitted \\[5pt]
    $\gamma_{\rm HARPS-N} [\ms]$ & $46932.9\pm1.7$ & - &  \multicolumn{2}{c}{  46543$\pm$1 }  & fitted \\	[5pt]
    $jitter_{\rm HDS} [\ms] $& $< 1.2$ & - & - & - & fitted \\[5pt]
    $jitter_{\rm HIRES} [\ms]$ & $7.0_{-2.0}^{+2.9}$ & - & - & - & fitted \\[5pt]
    $jitter_{\rm HARPS-N} [\ms]$ & $2.6\pm0.7$  & - & \multicolumn{2}{c}{  $ 1.80 \pm 0.43$ } & fitted  \\[5pt]
	$\dot{\gamma}$ [$\ms \rm day^{-1}]$ & $0.0017\pm0.0018$  & - & \multicolumn{2}{c}{0.0531$\pm$0.0087 }  & fitted  \\ [5pt]
	$\ddot{\gamma}$ $[\ms \rm day^{-2}]$ & $-2.03\pm0.21 \cdot 10^{-5}$ & - & - & - & fitted \\[5pt]
    $M\sin i$  [M$_{Jup}$]   &   $0.597\pm0.021$\tablefootmark{a} & $ \ge 1.8 $   &   $0.259\pm0.014$  & $1.370\pm0.053$ & derived \\ [5pt]
	Radius $R ~[\Rjup]$  & $1.019\pm0.031$ & - & - & - & derived \\[5pt]
	Density $\rho$ [$\rm g\;cm^{-3}$] & $0.70_{-0.064}^{+0.072} $ & - & - & - & derived \\[5pt]
	Surface gravity $\log g$ [cgs] & $3.15\pm0.03$ & - & - & - & derived \\[5pt]
	$a$ [AU]  &  0.03673$\pm$0.00064 & - & $0.1344\pm0.0025$  & $0.4756\pm0.0087$  & derived \\\noalign{\smallskip}
	\hline
	\end{tabular} 
	\tablefoot{
	\tablefoottext{a}{For XO-2Nb this corresponds to the real mass of the planet. It was obtained with a Monte-Carlo analysis using the median values of the orbital period and inclination angle of the orbital plane listed in Table \ref{table:transitparam}, the stellar mass in Table \ref{table:stellarparam} (second value), and the amplitude K of the radial velocity included in this Table.}}
    \end{table*}


    \begin{table*}
    \caption{ Radial velocities (RV), bisector velocity span (BIS) and activity index $\log$(\textit{R$^{\prime}_{\rm HK}$}) of XO-2N, as measured with the HARPS-N spectrograph. The BIS reported here are those derived by the data reduction pipeline of HARPS-N, with uncertainties assumed to be twice those on the radial velocities.}
    \label{table:radvelxo2n}
	\centering
	\small
	\begin{tabular}{cccccc}
	\hline
    \noalign{\smallskip}
    Time  & RV & RV error & BIS & $\log$(\textit{R$^{\prime}_{\rm HK}$}) & error $\log$(\textit{R$^{\prime}_{\rm HK}$}) \\  
	(BJD UTC - 2,450,000) & (km s$^{-1}$) & (km s$^{-1}$) & (km s$^{-1}$) & &  \\ 
    \noalign{\smallskip}
    \hline
    \noalign{\smallskip}
	
    6252.773594 &  47.0169 &   0.0047 &   0.0209 & -5.172 &  0.157 \\
    6266.500236 &  46.9521 &   0.0018 &   0.0204 & -4.871 &  0.022 \\
    6266.511139 &  46.9500 &   0.0016 &   0.0213 & -4.870 &  0.019 \\
    6266.522077 &  46.9472 &   0.0019 &   0.0187 & -4.905 &  0.026 \\
    6266.533015 &  46.9416 &   0.0018 &   0.0164 & -4.907 &  0.025 \\
    6266.543953 &  46.9406 &   0.0021 &   0.0210 & -4.927 &  0.033 \\
    6266.554891 &  46.9490 &   0.0028 &   0.0303 & -4.874 &  0.043 \\
    6266.565829 &  46.9422 &   0.0024 &   0.0245 & -4.891 &  0.035 \\
    6266.576768 &  46.9413 &   0.0028 &   0.0249 & -4.986 &  0.054 \\
    6266.587706 &  46.9388 &   0.0023 &   0.0150 & -4.889 &  0.034 \\
    6266.598644 &  46.9305 &   0.0023 &   0.0229 & -4.855 &  0.031 \\
    6266.609593 &  46.9217 &   0.0022 &   0.0274 & -4.881 &  0.030 \\
    6266.620520 &  46.9181 &   0.0018 &   0.0195 & -4.889 &  0.023 \\
    6266.631458 &  46.9153 &   0.0017 &   0.0205 & -4.910 &  0.023 \\
    6266.642396 &  46.9162 &   0.0016 &   0.0178 & -4.901 &  0.019 \\
    6266.653334 &  46.9196 &   0.0015 &   0.0209 & -4.872 &  0.017 \\
    6266.664284 &  46.9179 &   0.0016 &   0.0217 & -4.901 &  0.020 \\
    6266.675210 &  46.9135 &   0.0015 &   0.0191 & -4.890 &  0.017 \\
    6266.686148 &  46.9138 &   0.0015 &   0.0152 & -4.877 &  0.017 \\
    6297.623295 &  46.9958 &   0.0015 &   0.0154 & -4.928 &  0.019 \\
    6298.643938 &  46.8409 &   0.0022 &   0.0097 & -4.921 &  0.034 \\
    6299.762797 &  47.0128 &   0.0015 &   0.0099 & -4.907 &  0.020 \\
    6305.631219 &  46.9708 &   0.0020 &   0.0125 & -4.938 &  0.031 \\
    6322.563804 &  46.8777 &   0.0015 &   0.0129 & -4.898 &  0.019 \\
    6323.686453 &  47.0147 &   0.0033 &   0.0209 & -4.947 &  0.072 \\
    6380.349341 &  46.9211 &   0.0033 &   0.0106 & -5.269 &  0.155 \\
    6400.444868 &  46.8469 &   0.0100 &   0.0573 & -4.861 &  0.214 \\
    6406.361875 &  46.8938 &   0.0036 &   0.0113 & -4.908 &  0.060 \\
    6420.392578 &  47.0089 &   0.0031 &   0.0066 & -5.229 &  0.114 \\
    6437.372927 &  46.8412 &   0.0014 &   0.0140 & -4.964 &  0.018 \\
    6548.731988 &  47.0043 &   0.0024 &   0.0059 & -4.949 &  0.044 \\
    6586.684893 &  46.8605 &   0.0024 &   0.0145 & -4.989 &  0.044 \\
    6602.751791 &  46.9259 &   0.0025 &   0.0094 & -5.042 &  0.055 \\
    6616.709751 &  47.0041 &   0.0027 &   0.0203 & -4.874 &  0.042  \\
    6617.722042 &  46.8368 &   0.0017 &   0.0153 & -4.929 &  0.023 \\
    6631.704304 &  46.9647 &   0.0020 &   0.0051 & -5.035 &  0.041 \\
    6657.453439 &  46.8928 &   0.0056 &   0.0255 & -5.086 &  0.183 \\
    6700.404020 &  47.0066 &   0.0026 &   0.0147 & -4.983 &  0.051 \\
    6701.349611 &  46.8410 &   0.0019 &   0.0164 & -4.963 &  0.031 \\
    6783.385415 &  46.9513 &   0.0014 &   0.0077 & -4.996 &  0.024 \\
    6784.360848 &  46.9543 &   0.0016 &   0.0082 & -4.980 &  0.028 \\
    6933.733982 &  46.9031 &   0.0023 &   0.0052 & -5.009 &  0.050 \\
    6935.754527 &  47.0090 &   0.0017 &   0.0107 & -4.996 &	0.030 \\
    \noalign{\smallskip}
    \hline
 \end{tabular} 
 \end{table*}


	\subsection{Analysis of the radial velocity time series}
\label{sec:radveltimeseries}
 To refine the orbital parameters of XO-2N\,b, our new HARPS-N data were analysed along with the RVs obtained with the spectrographs HDS (High Dispersion Spectrograph) at the Subaru telescope \citep{narita11}, and HIRES (High Resolution Echelle Spectrograph) at the Keck telescope \citep{knutson14}. The original spectroscopic data from \cite{burke07} were not included because they have considerably larger error bars  ($> 20 \ms$) than the other datasets and thus  do not yield any significant improvement of the orbital solution. 
 We first fitted separately the HDS, HIRES, and HARPS-N RV time series with a Keplerian orbit and a slope, after removing the in-transit points which are affected by the Rossiter-McLaughlin effect. We then performed a Bayesian analysis with our Differential Evolution Markov Chain Monte Carlo (DE-MCMC) tool (see, e.g., \citealt{desidera14}). \citet{narita11} and \citet{knutson14} reported on a positive acceleration revealed in their RV datasets, equal to $0.0206\pm0.0016$ and $0.0126_{-0.0036}^{+0.0039}~\ms \rm day^{-1}$ respectively, possibly due to a second massive companion in an outer orbit, namely XO-2N\,c. Our observations and analysis confirm the existence of this trend, but we observed a turn-over in the HARPS-N data and measured a negative acceleration equal to $-0.0170_{-0.0031}^{+0.0033}~\ms \rm day^{-1}$. 
 
 We then modelled the RV data of the three datasets with a Keplerian orbit and a polynomial model following the formalism of \citet{kipping11} and choosing as \textit{$t_{\rm pivot}$} in their Eq. (1) the mean of the epochs of the RV observations. In total, our RV model has thirteen free parameters: 
the transit epoch \textit{T$\rm _c$}, the orbital period \textit{P}, the radial-velocity semi-amplitude \textit{K}, \textit{$e\,\cos\omega$} and \textit{$e\,\sin\omega$} of XO-2N\,b, where \textit{e} and \textit{$\omega$} are the eccentricity and argument of periastron; three systemic velocities (\textit{$\gamma_{\rm HDS}$}, \textit{$\gamma_{\rm HIRES}$}, \textit{$\gamma_{\rm HARPS-N}$}) and RV jitter terms ($jitter_{\rm HDS}$, $jitter_{\rm HIRES}$, $jitter_{\rm HARPS-N}$), corresponding to the three different datasets; the linear $\dot{\gamma}$ and quadratic $\ddot{\gamma}$ trends. Gaussian priors were imposed on i) \textit{T$\rm _c$} and \textit{P} from the new ephemeris we derived in Sect.\ref{sec:lcanalysis}, and ii) the centre times of the secondary eclipse observed by \citet{machalek09} at 3.6, 4.5, 5.8, and 8.0 $\mu$m with Spitzer,
because these provide strong constraints on the orbital eccentricity; for this purpose, we used Eq. (18) in \citet{jordan08}. We adopted Jeffrey's priors for the three jitter terms (see, e.g., \citealt{gregory05}) and non-informative priors for the RV semi-amplitude and the three systemic velocities.
 
\begin{figure}
   \centering
   \includegraphics[width=0.4\textwidth, angle =90]{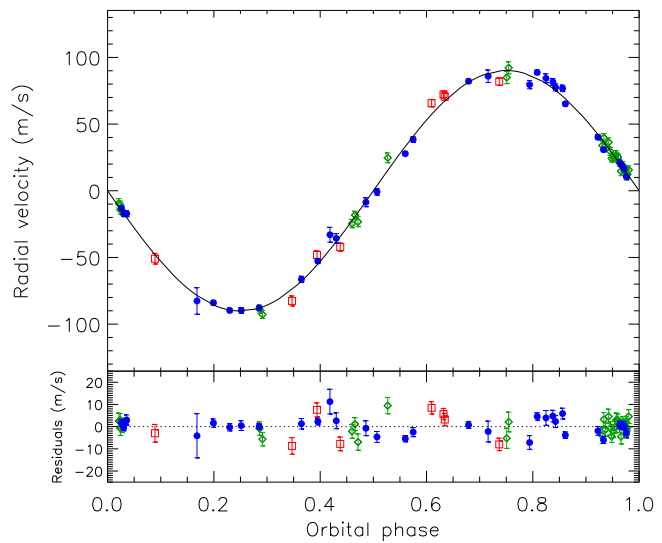}
   \smallskip
   \caption{Phase-folded radial-velocity curve of XO-2N\,b, after removing the quadratic trend.
Green diamonds, red squares, and blue circles indicate the HDS, HIRES, and HARPS-N
measurements respectively. The black solid line shows the best fit model.}
   \label{fig:rvxo2nphase}
   \end{figure}

 After the earliest analysis, we noticed that the first HIRES point at 2454372.14~$\rm BJD_{UTC}$ was $3\sigma$ below our best fit hence we discarded it and repeated the analysis, although its inclusion has a very minor influence on the final solution. The posterior distributions of the fitted and derived parameters were obtained with the same procedure as described in \cite{desidera14}. Their medians and $68.3\%$ confidence intervals are reported in Table~\ref{table:rvxoharpsn}, along with the system parameters of XO-2S derived by \cite{desidera14}. The Keplerian best fit to the three datasets, after removing the quadratic trend, is shown in Fig. \ref{fig:rvxo2nphase}. Our new RV semi-amplitude $K=89.78 \pm 0.81 \ms$ is slightly lower than the value found by \citet{knutson14}, that is $K=93.9 \pm 2.2 \ms$. Its precision is improved by a factor of three, especially thanks to the HARPS-N data. The orbital eccentricity is consistent with zero, $e < 0.006$ at $1\sigma$, which justifies the use of a circular model for the transit modelling (Sect.\ref{sec:lcanalysis2}). The combined analysis of the complete RV dataset and of the new transit light curves allows us to refine the physical parameters of the planet XO-2N\,b, and we find that the best values for the radius, mass and density are $1.019\pm0.031\Rjup$, $0.597\pm0.021\Mjup$, and $0.70_{-0.064}^{+0.072} \rm g\;cm^{-3}$ respectively.

 	\subsubsection{RV acceleration in XO-2N: evidence for an outer companion or solar-like stellar activity cycle?}
\label{sec:radvelacceleration}

\begin{figure}
   \centering
   \includegraphics[width=0.45\textwidth]{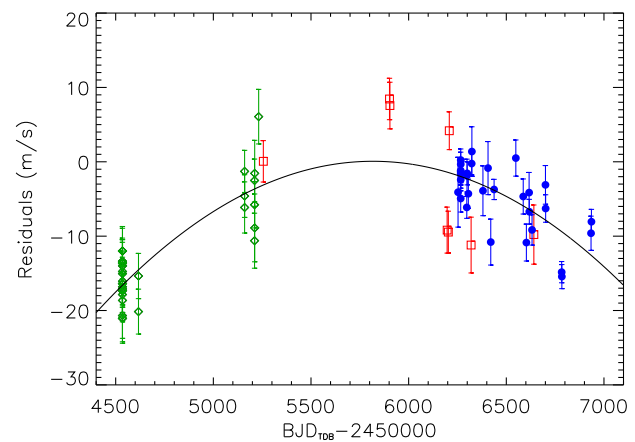}
   \smallskip
   \caption{Long-term trend found in the RV residuals data of XO-2N after removing the orbital signal of XO-2N\,b, caused either by a long-period companion XO-2N\,c or to the activity cycle of the star. Green diamonds, red squares, and blue circles indicate the HDS, HIRES, and HARPS-N
measurements respectively, while the black solid line displays the best fit of the quadratic trend. Residuals of RV measurements with an error bar greater than 5 m s$^{-1}$, even though considered in the analysis, were not included in this plot for a better visual appreciation the long-term trend. }

   \label{fig:rvxo2nresiduals}
   \end{figure}

 Under the hypothesis that an outer massive companion exists and is responsible for the observed acceleration, the change of sign derived from HARPS-N data with respect to the previous measurements could be likely caused by the curvature of the orbit of XO-2N\,c. Figure \ref{fig:rvxo2nresiduals} displays the quadratic trend in the RV residuals, after removing the orbital signal of XO-2N\,b and considering all the available datasets. By assuming a circular orbit for the outer companion XO-2N\,c, the derived quadratic trend allows us to estimate its orbital period and minimum mass (see \citealt{kipping11}): $P\ge17$~yrs and $M\sin{i}\ge1.8~\Mjup$. Direct imaging observations collected with the AstraLux instrument did not reveal a companion to XO-2N with $\Delta$\textit{i}$^{\prime}$ $\lesssim$ 8 magnitudes and a separation $\lesssim$ 2$^{\prime\prime}$ (\citealt{daemgen09}), which corresponds to $\sim$300 AU at the star location. A rough estimate for an upper limit of the XO-2N\,c mass can be made using the information on $\Delta$\textit{i}$^{\prime}$. By fixing the age and the metallicity to the values of XO-2N, and using the CMD\footnote{http://stev.oapd.inaf.it/cgi-bin/cmd} stellar isochrones, we find that the unseen companion should have a mass M$\lesssim$0.16 M$_\odot$, which roughly corresponds to a main-sequence star of spectral type later than M5V-M6V. 
 
\begin{figure}
   \centering
   \includegraphics[width=9cm]{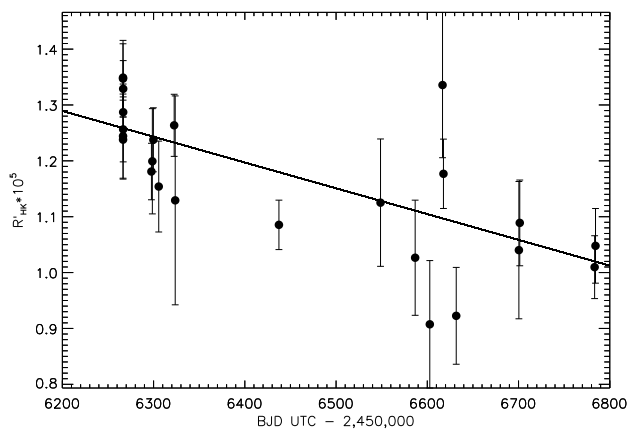}
   \smallskip
   \caption{Time series of the chromospheric activity index \textit{R$^{\prime}_{\rm HK}$} for XO-2N extracted from the HARPS-N spectra. Values of log(\textit{R$^{\prime}_{\rm HK}$}) are listed in Table \ref{table:radvelxo2n}. We considered only the spectra with a signal-to-noise greater than 4 in the 6$^{th}$ order. Superposed is the best-fit linear function, revealing the presence of a negative trend.}
   \label{fig:rhktimeseries}
   \end{figure}

\begin{figure}
   \centering
   \includegraphics[width=9cm]{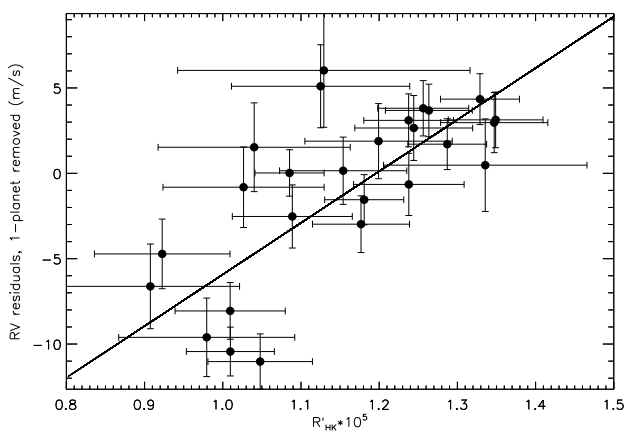}
   \smallskip
   \caption{Correlation between the RV residuals and the chromospheric activity index \textit{R$^{\prime}_{\rm HK}$} for XO-2N. Data are extracted from the HARPS-N spectra. Superposed is the best-fit linear function, showing the existence of a positive trend.}
   \label{fig:rhkrvrescorr}
   \end{figure}

 If not due to an outer companion, the quadratic trend could reflect a solar-like activity cycle, which leaves an imprint in the RV measurements over a timespan of several years. This is consistent with the observation that older stars, with longer rotation periods, have longer activity cycles (e.g., \citealt{noyes84}; \citealt{ossendri97}). In their work, \cite{knutson14} do not mention for XO-2N the existence of a correlation between their RV residuals and some indicators of the stellar activity, as \textit{ R$^{\prime}_{\rm HK}$} or \textit{S$_{\rm HK}$}, which would call the planetary origin of the acceleration into question. The same situation does not occur in our dataset. The time series of the \textit{R$^{\prime}_{\rm HK}$} activity index, calculated according to the recipe described in \cite{lovis11}, shows the existence of a negative trend (see Fig. \ref{fig:rhktimeseries}). We modelled the data with a linear fit, obtaining a negative slope with a 1$\sigma$ uncertainty of $\sim14\%$. The Spearman's rank correlation coefficient of the data is quite high ($\rho$=-0.79) and with a low FAP (1.3$\cdot10^{-6}$)\footnote{Values obtained with the \texttt{R$\char`_$CORRELATE} function in IDL.}, thus strengthening the real existence of a correlation. The direction of the trend being similar for the RV residuals and \textit{R$^{\prime}_{\rm HK}$}, a significant positive correlation is observed between them (see Fig. \ref{fig:rhkrvrescorr}), characterised by a Spearman's rank correlation coefficient $\rho$=+0.68 (FAP = 1$\cdot10^{-4}$) for N=26 measurements. 
 
Due to the importance of this result, the significance of the correlation was further investigated by applying two different tests. Through the first method, described in \cite{figueira14} (and references therein) to which we refer to as the \textit{null hypothesis} approach, we evaluated the probability of having a Spearman's rank correlation coefficient larger or equal than that of the real data by testing the null hypothesis, i.e. by assuming that the data pairs are actually uncorrelated. This method, which does not take into account the data uncertainties, employs a Fisher-Yates data shuffling to generate fake data pairs which are uncorrelated. We generated 10,000 such random datasets and then calculated the z-score for the correlation coefficient of the original dataset relative to the distribution of the simulated Spearman's rank correlation coefficients. The corresponding probability (the p-value) is calculated as the one-sided likelihood of having such a z-score from the observed Gaussian distribution. The second method, which instead takes into account the data uncertainties, is described in \cite{curran14} (referred to as \textit{composite} method) and it employs a double Monte-Carlo re-sampling of the dataset pairs. First, fake data pairs are generated through a bootstrap (with re-sampling) method, with new data pairs randomly chosen from the original ones (i.e. each new pair comes from the real dataset and it may appear more than once or not at all in a simulated dataset). Then, the second step consists in randomly shuffling each measurement within its uncertainty following a normal distribution of width 1 and centred on 0. These two steps are repeated 10,000 times, resulting in a distribution of Spearman's rank correlation coefficients for which the mean and the standard deviation are calculated. Moreover, the z-score is calculated for each simulated dataset through the Fisher transformation of the correlation coefficient, as discussed in \cite{curran14}, resulting in a z-score distribution for which the mean and the standard deviation are provided. The z-score defines the level of significance of a correlation, being the number of $\sigma$ from a null correlation for the original dataset. 

The null hypothesis test applied to the RV residuals and \textit{R$^{\prime}_{\rm HK}$} dataset for XO-2N resulted in z-score = -3.41, corresponding to p-value = 0.032$\%$, while the composite method resulted in $\rho_{Spearman}$=0.54$\pm$0.15 and z-score = 1.267$\pm$0.451. Both results suggest that the null-hypothesis can be rejected for our dataset. 

A linear regression applied to the same dataset results in a slope of 30.2 \ms per unit of \textit{R$^{\prime}_{\rm HK}$}, with a 1$\sigma$ uncertainty of $\sim10\%$. Comparing this slope value with that predicted by the empirical model described by Eq. (13) in \cite{lovis11} for an inactive star with the same $T_{\rm eff}$ and [Fe/H] of XO-2N, i.e. 32.9$\pm$0.2 \ms per unit of \textit{R$^{\prime}_{\rm HK}$}, we find that our estimate is in agreement with that expected for RV variations produced by a long-term stellar activity cycle.
 
 The positive correlation found between the RV residuals and \textit{ R$^{\prime}_{\rm HK}$} indicates that the variations seen in the RV residuals are likely related to the intrinsic behaviour of the star. In fact, during a phase of high stellar activity a net redshift should appear in the RV measurements as a consequence of the suppressed convective motion in the photospheric active regions, which gives a blueshift contribution (see, for instance, \citealt{dumusque12}, and references therein). This means that a negative trend in the RVs is expected when a star is observed during a declining phase of its stellar cycle, with a similar trend appearing in the \textit{R$^{\prime}_{\rm HK}$} index.
 Despite in our data there is evidence in favour of the stellar activity cycle as the cause of the RV variations over nearly 2 years, in our opinion the covered timespan is not yet sufficiently long to corroborate this conclusion and to exclude the outer massive companion as an alternative explanation. More data  are indeed necessary over a longer time baseline to cover at least one activity cycle of XO-2N and to investigate in greater detail the correlation with the RV measurements.
 
	\subsection{The Rossiter-McLaughlin effect}
	\label{sec:rml}
\begin{figure}
   \centering
   \includegraphics[width=8cm]{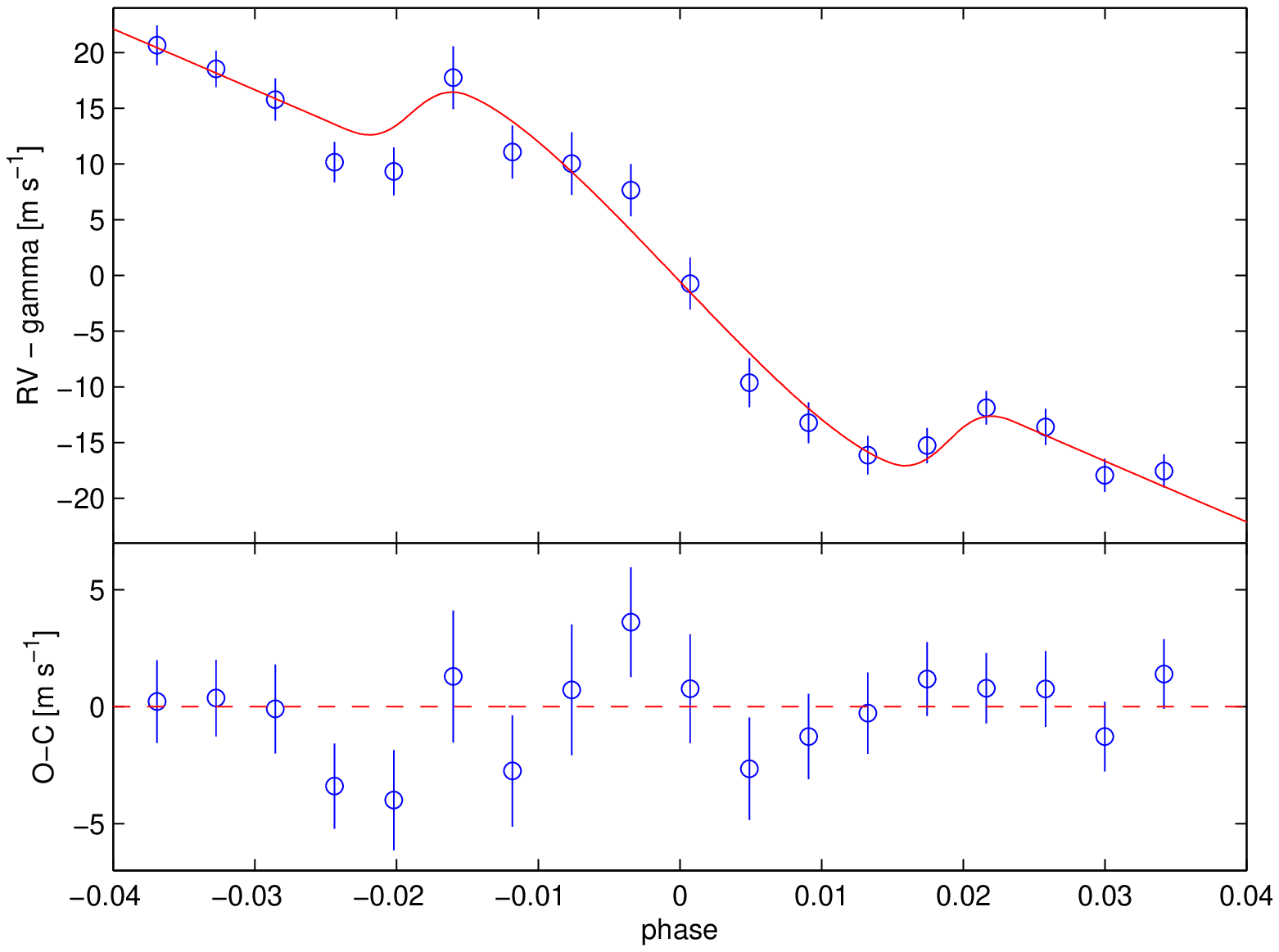}
   \includegraphics[width=8cm]{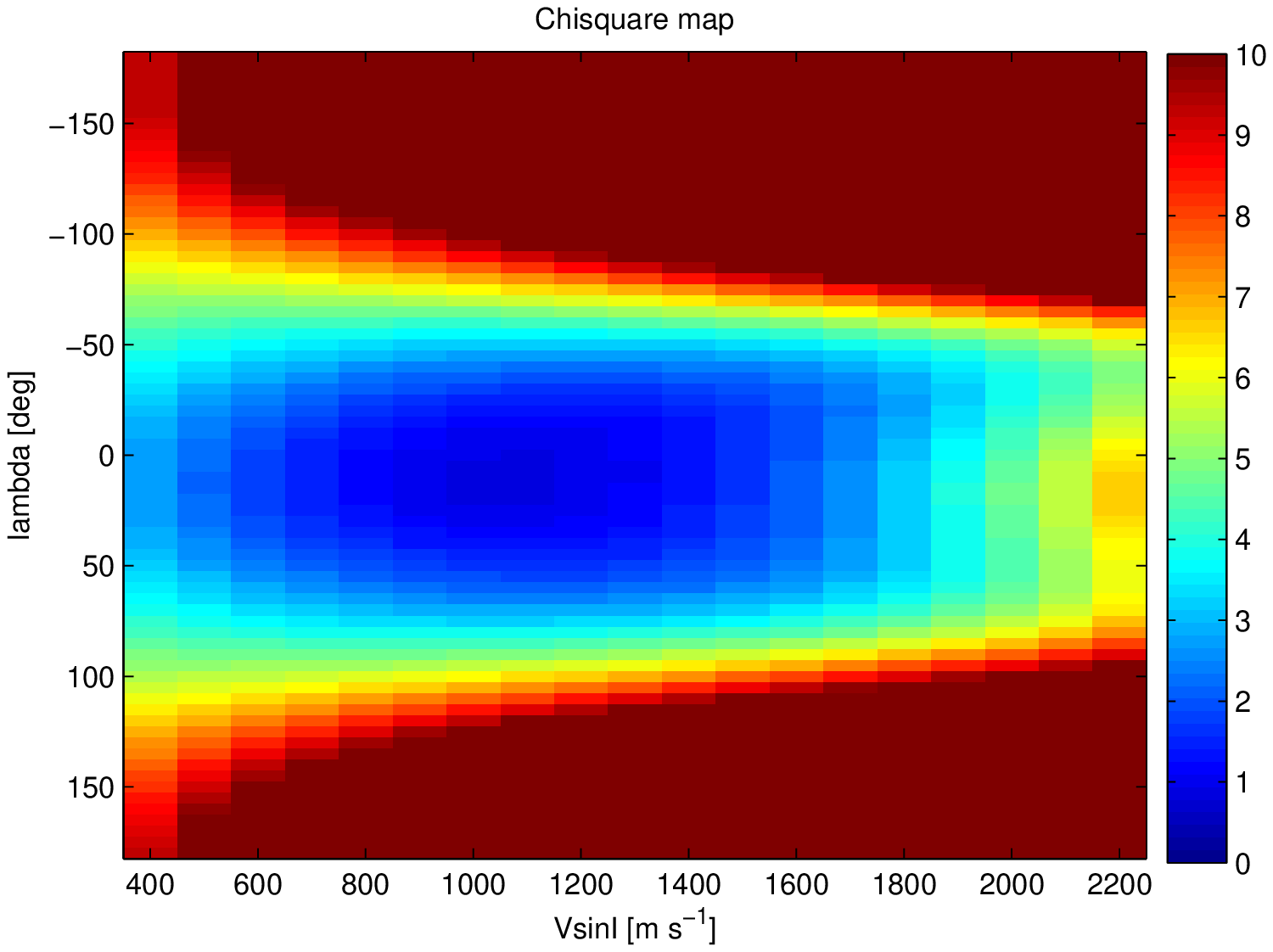}
   \caption{\textit{Upper panel}. Radial velocity variations of the XO-2N star in correspondence to the transit of the exoplanet XO-2Nb occurred on the night between 4 and 5 December, 2012. Data are folded according to the transit ephemeris derived in this study [T$_{c}$=2,455,565.546480 + N*2.6158592 BJD TDB], and superposed is our best-fit RML model (red line). \textit{Lower panel}. The $\chi^{2}$ map in the $\lambda$ - \textit{V}sin\textit{I$_{\star}$} parameter space.}
   \label{fig:RMLdata}%
    \end{figure}

    Dedicated observations of the RML effect for the XO-2N system were carried out with HARPS-N during the transit which occurred in 2012, December 4-5. The RML observations for this system were previously performed by \cite{narita11}, who concluded that the orbit of XO-2Nb is unlikely to be highly tilted, but they obtained a poorly determined estimate for the projected spin-orbit angle, \textit{$\lambda$} = 10$^{\circ}\pm$72$^{\circ}$. Here we present updated results obtained from the analysis of 18 spectra collected at that epoch, starting $\sim$1 hour before the beginning of the transit up to $\sim$1 hour after the last contact. The RML effect was modelled with the fitting algorithm described in \cite{covino13} and \cite{esposito14}. Figure \ref{fig:RMLdata} shows the RV measurements used for the analysis of the RML effect, folded according to the ephemeris calculated in this work, with the best-fit superposed. We found a projected spin-orbit angle \textit{$\lambda$} = 7$^{\circ}\pm$11$^{\circ}$, and derived for the projected rotational velocity of the host star \textit{V}sin\textit{I$_{\star}$} = 1.07$\pm$0.09 km/s. The $\chi^{2}$ map in the $\lambda$ - \textit{V}sin\textit{I$_{\star}$} parameter space is shown in the same figure. The uncertainties were derived with a Monte-Carlo analysis taking into account the uncertainties of all the parameters used. Our result provides a stronger evidence for a spin-orbit alignment with an uncertainty $\sim$7 times lower. In fact, our new estimate of the transit impact parameter partially breaks the existing degeneracy between $\lambda$ and the stellar projected rotational velocity \textit{V}sin\textit{I$_{\star}$}, resulting in a better constraint on $\lambda$. 
    
    \subsection{The XO-2N spin-orbit alignment in three-dimensions}
	\label{sec:alangle}
    Thanks to the photometric measurement of the XO-2N rotation period, together with the analysis of transit photometry and the RML effect, we have the opportunity to provide an estimate of the angle \textit{$\psi$} between the stellar rotation axis and the normal to the planetary orbital plane. This is given by the formula \textit{$\cos\psi$}=cos\textit{I$_{\rm \star}$}cos\textit{i}+sin\textit{I$_{\star}$}sin\textit{i}cos\textit{$\lambda$} (see discussion in Sect.4.4 of \citealt{winn07}), where \textit{I$_{\rm \star}$} is still the inclination between the stellar spin axis and the direction of the line of sight, \textit{i} is the orbital inclination angle as measured by transit photometry, and \textit{$\lambda$} is the projected spin-orbit angle derived from the analysis of the RML effect. By calculating \textit{$\sin$I$_{\rm \star}$} from the projected rotational velocity of the star \textit{V$\sin$I$_{\rm \star}$}, the rotation period (41.6$\pm$1.1 days) and the more precise measurement of the stellar radius we have derived (0.998$\pm$0.033 R$_\odot$), we obtain \textit{$\sin$I$_{\rm \star}$}=0.88$\pm$0.09. By propagating the errors, we finally obtain \textit{$\cos\psi$}=0.89$\pm$0.12, as expected very similar to \textit{$\sin$I$_{\star}$} because the angles \textit{i} and \textit{$\lambda$} are close to 90$^{\circ}$ and 0$^{\circ}$. By neglecting the unphysical values for which \textit{$\cos\psi$}>1, this result corresponds to \textit{$\psi$}=27$^{+12}_{-27}$ degrees. Because a measurement of the angle \textit{$\psi$} is difficult to make, out of the 88 planet host stars for which $\lambda$ was determined, only for nine of them \textit{$\psi$} was constrained \footnote{http://www.astro.keele.ac.uk/jkt/tepcat/rossiter.html}. For XO-2N, our estimate is the first available.

\section{Stellar activity}
\label{sec:stellaractivity}

In this work we characterise the stellar activity of the XO-2 companions, both through photometric measurements and spectroscopic analysis.  The analysis of the photometric time series resulted in an estimate of the spin rotation periods for the two components (Sect.\ref{sec:lcanalysis}), we then used the HARPS-N spectra to calculate the chromospheric activity index \textit{R$^{\prime}_{\rm HK}$} based on the analysis of the \ion{Ca}{ii} H$\&$K lines. Moreover, an analysis of different CCF line shape parameters (asymmetry indicators) was carried out to look for periodical modulations and correlations with the radial velocities and the \textit{R$^{\prime}_{\rm HK}$} index. 

\cite{figueira13} introduced and compared several indicators for the asymmetry of the CCF and shown how they can be used to establish whether the radial velocity variation of a star is due to line profile distortions induced by, e.g., stellar activity. We have computed the indicators bisector inverse span (BIS)\footnote{We have verified that between the line-bisector calculated by the DRS and the BIS indicator calculated with the method of \cite{figueira13} there is a very high correlation (R=0.99) for both stars, then we decided to adopt the latter in our discussion.}, radial velocity difference \textit{$\Delta$V}, and velocity asymmetry \textit{V$_{asy}$} for each of the XO-2 components, and a comparison between those parameters and the activity index was also performed. We adopted procedures similar to those of \cite{figueira13} and \cite{santos14} to compute the asymmetry parameters with the exception of $V_{\rm asy}$ for which the derivative of the CCF with respect to the RV was numerically evaluated by means of a Savitzky-Golay filter (e.g., \citealt{press92}). We estimated the standard deviation of the indicator $\Delta$\textit{V} from the standard deviations of the best fit RVs as obtained with the Levenberg-Marquardt IDL procedure {\tt MPFIT.PRO} applied to fit a Gaussian and a bi-Gaussian profiles to the CCF, respectively. The standard deviation of $V_{\rm asy}$ was computed with a bootstrap procedure by taking into account both the photon shot noise and the correlated noise of the CCF.

	\begin{table}
	\scriptsize
	\centering
	\caption{Summary of stellar activity-related information for the XO-2 system. The values of the activity index log(\textit{R}$^{\prime}_{\rm HK}$) are given as weighted mean, and the uncertainty of each parameter is the standard deviation of the mean. The estimates for the stellar rotation periods correspond to two different relations, those of \cite{noyes84} [N84] and \cite{mamajek08} [M08].}   
	 \label{table:actindtable}
	\begin{tabular}{ccc}
	\hline
    \noalign{\smallskip}
     Parameter		 & XO-2N & XO-2S \\ 
		 				& (43 spectra) & (63 spectra) \\
      \noalign{\smallskip}
      \hline
      \noalign{\smallskip}
	 	
    $\log$ (R$^{\prime}_{\rm HK}$) &  -4.91$\pm$0.01  & -5.02$\pm$0.01 \\ [5pt]
    Exp. stellar rot. period [days]  & 39.6$\pm$0.6 [N84] & 41.1$\pm$0.7 [N84] \\ [5pt]
     & 41.8$\pm$0.8 [M08]  & 44.7$\pm$0.7 [M08] \\ [5pt] 
    Meas. stellar rot. period [days]  & 41.6$\pm$1.1 & 26.0$\pm$0.6 (or 34.5 days)\\
    \noalign{\smallskip}
    \hline 
	\end{tabular}
	\end{table}	


   \begin{figure*}
   \centering
   \includegraphics[width=18cm]{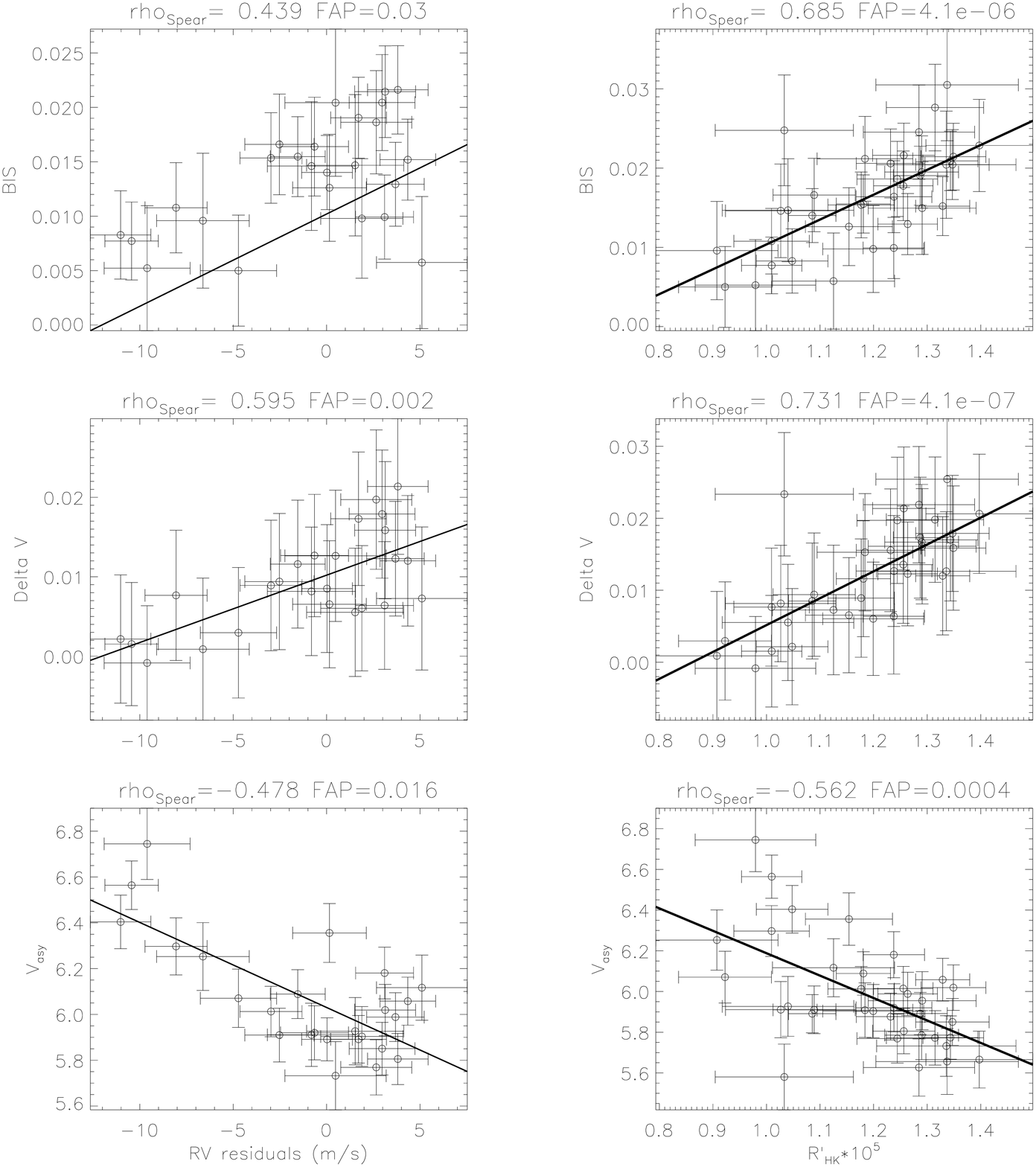}
   \caption{Correlations between the asymmetry indicators and the residuals of the radial velocities (\textit{left panel}) for the star XO-2N, and between the same indicators and the activity index \textit{R$^{\prime}_{\rm HK}$} (\textit{right pane}l). We considered only values of \textit{R$^{\prime}_{\rm HK}$} extracted from spectra with a signal-to-noise greater than 4 in the 6$^{th}$ order.}
   \label{fig:asyindN}%
    \end{figure*}    
    
   \begin{figure}
   \centering
   \includegraphics[width=9.5cm]{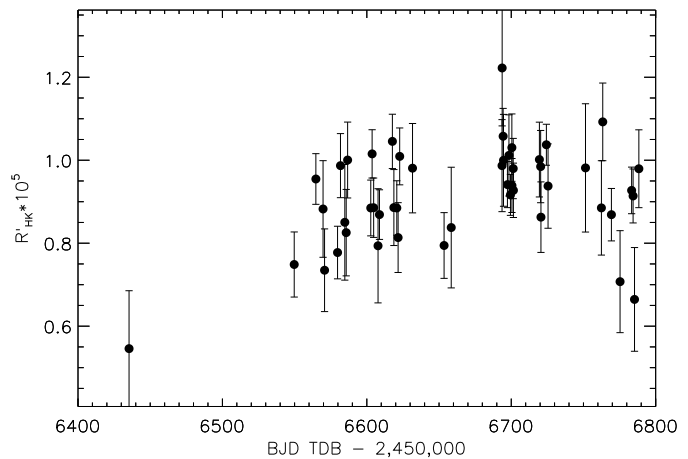}
   \caption{Time series of the chromospheric activity index \textit{R$^{\prime}_{\rm HK}$} for the star XO-2S. Only values of \textit{R$^{\prime}_{\rm HK}$} derived from spectra with S/N > 3.9 in the 6$^{th}$ order were considered.}
   \label{fig:rhktimexo2s}%
   \end{figure} 
    
   \begin{figure*}
   \centering
   \includegraphics[width=18cm]{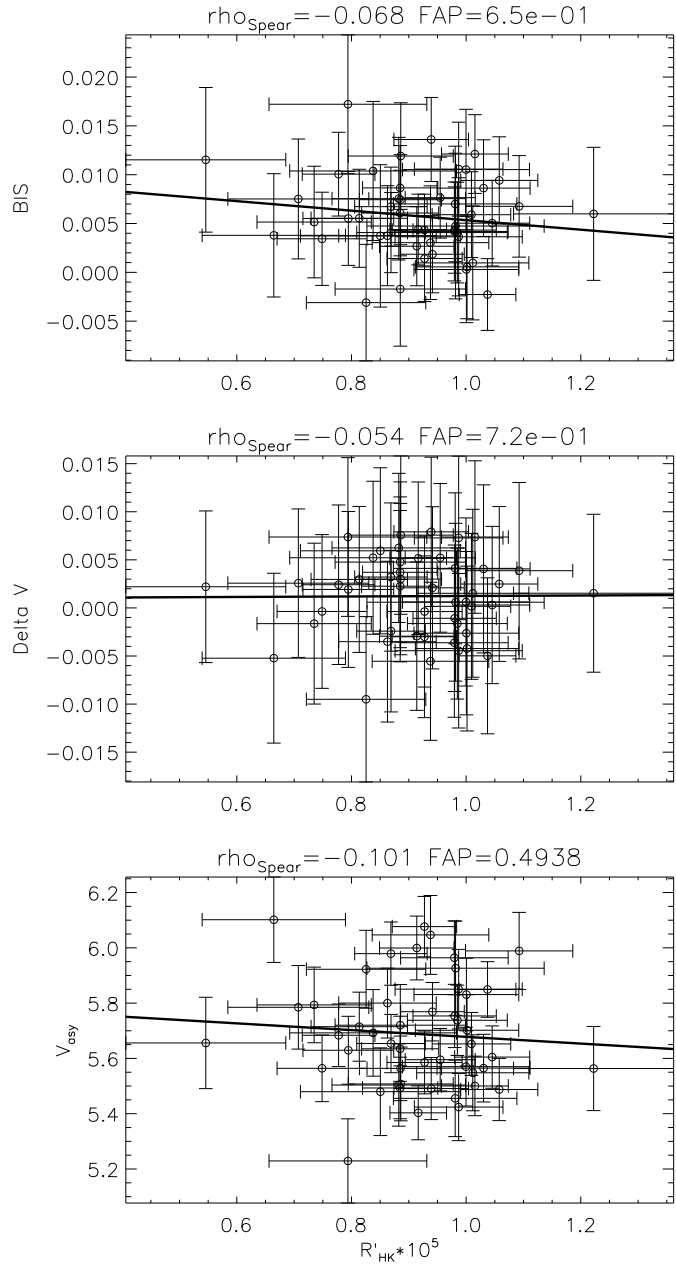}
   \caption{Correlations between the asymmetry indicators and the chromospheric activity index \textit{R$^{\prime}_{\rm HK}$} for the star XO-2S. Only values of \textit{R$^{\prime}_{\rm HK}$} derived from spectra with S/N > 3.9 in the 6$^{th}$ order were considered for the analysis.}
   \label{fig:asyindS}%
   \end{figure*}      
         
   \begin{figure}
   \centering
   \includegraphics[width=0.55\textwidth]{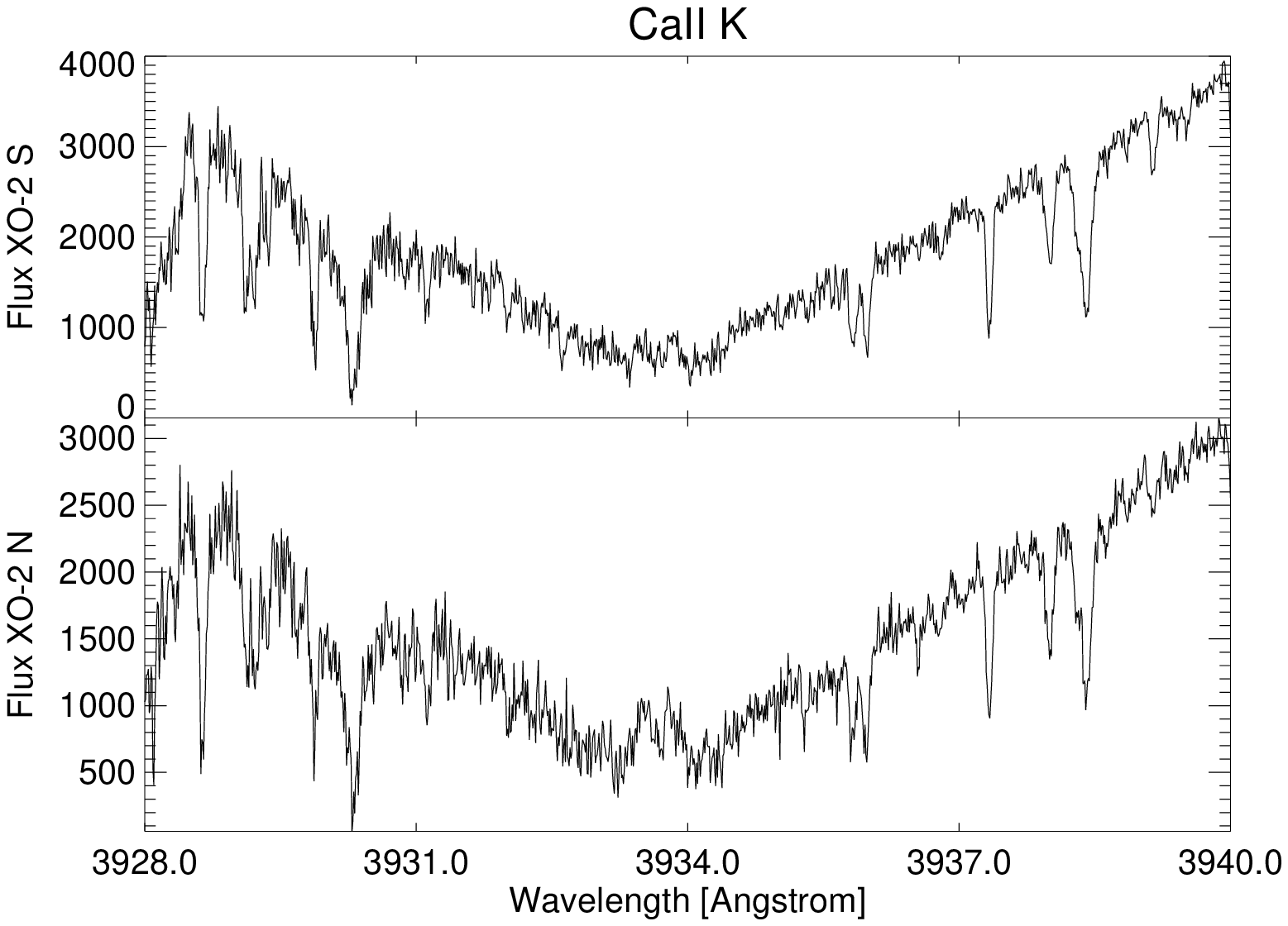}
   \includegraphics[width=0.55\textwidth]{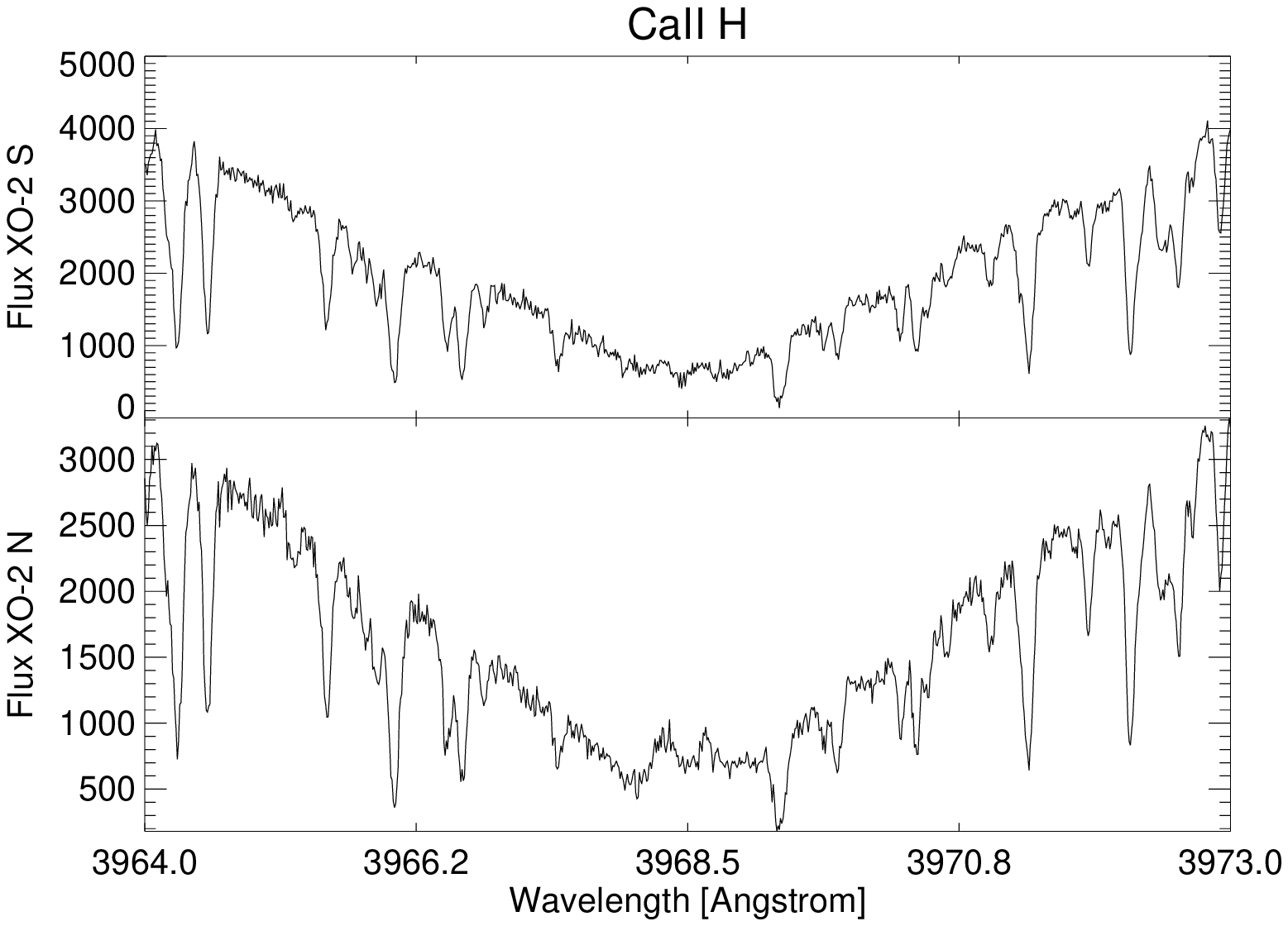}
   \caption{\ion{Ca}{ii} K (upper panel) and H (lower panel) lines of the coadded HARPS-N spectra of XO-2N and XO-2S. It is clearly visible a greater emission flux in the line cores of the North component, which is indicative of a greater activity of this star.}
   \label{fig:coaddhk}%
    \end{figure}


	\subsection{XO-2N}
	\label{sec:stellaractivity1}
    The analysis of the \textit{R$^{\prime}_{\rm HK}$} index time series has been presented in Sect.\ref{sec:radveltimeseries}, where we discussed the correlation found with the RV residuals, possibly indicating that \textit{R$^{\prime}_{\rm HK}$} is outlining the activity cycle of the star.
     
    Here we investigate the correlations between the asymmetry indicators and the residuals of the radial velocities (i.e. removing the one-planet fit and the quadratic term), and the results are shown in the left panel of Fig. \ref{fig:asyindN}, while the right panel shows the correlations  between the same indicators and the activity index \textit{R$^{\prime}_{\rm HK}$}. We note that between the CCF asymmetry indicator $\Delta$\textit{V} and the RV residuals there is a moderate and low-FAP correlation, as quantified by the Spearman's rank correlation coefficients indicated on the top of each plot. More significant correlations are observed when considering the \textit{R$^{\prime}_{\rm HK}$} index instead of the RV residuals, in particular for the bisector BIS and $\Delta$\textit{V}. These seem to indicate that the activity of the star does influence the spectral line profile at some extent.  
    
    To better characterize the significance of these correlations, we performed the same analysis as described in Section \ref{sec:radvelacceleration} and based on two different tests. The results are summarized in Table \ref{table:signcorrcoeff}. They show that the null hypothesis can be discarded for substantially all the pairs, assuming $\sim$1$\%$ as the threshold for the p-value below which a correlation may be assumed significant. The analysis with the composite method suggests that three most significant correlations found by simply computing the Spearman's rank coefficient with the actual datasets, i.e. those between the RV residuals and the BIS/$\Delta$\textit{V} asymmetry indicators of the CCF, are likely less pronounced, given the limited number of measurements (N=25). Nonetheless, our analysis suggests that effects of the stellar activity are influencing the spectral line profiles.
   
We finally analysed the time series of the CCF asymmetry indicators with the GLS algorithm, to search for signals related to the stellar rotation period of $\sim$41 days measured via photometry. No significant peaks were found, and in particular no periodicity related to the star rotation appears for \textit{R$^{\prime}_{\rm HK}$}. 
	
	\begin{table*}
	\small
	\centering
	\caption{Results of the analysis carried out to asses the significance of the Spearman's rank correlations coefficients for the pairs of XO-2N datasets shown in Fig. \ref{fig:asyindN}.}   
	 \label{table:signcorrcoeff}
	\begin{tabular}{ccccccc}
	\hline
    \noalign{\smallskip}
     	Method	 & RV resid. \textit{vs} BIS & RV resid. \textit{vs} $\Delta$V & RV resid. \textit{vs} V$_{asy}$ &  R$^{\prime}_{\rm HK}$ \textit{vs} BIS & R$^{\prime}_{\rm HK}$ \textit{vs} $\Delta$V & R$^{\prime}_{\rm HK}$ \textit{vs} V$_{asy}$\\ 
      \noalign{\smallskip}
      \hline
      \noalign{\smallskip}	 	
    \textit{Null hypothesis} &  z-score = -2.14 & z-score = -2.93 & z-score = +2.36 & z-score = -4.04 & z-score = -4.32 & z-score = +3.31 \\ [3pt]
    &	p-value = 1.6$\%$ &	p-value = 0.9$\%$ &	p-value = 0.17$\%$ &	p-value = 0.003$\%$ &	 p-value = 0.05$\%$ &	p-value < 0.002$\%$ \\ [5pt]
    \textit{Composite Monte-Carlo} &  $\rho$ = +0.32$\pm$0.19  & $\rho$ = +0.34$\pm$0.18 & $\rho$ = -0.43$\pm$0.19 & $\rho$ = +0.44$\pm$0.14 & $\rho$ = +0.39$\pm$0.15 & $\rho$ = -0.40$\pm$0.15 \\ 
    \noalign{\smallskip}
    \hline 
    \noalign{\smallskip}
    Original $\rho_{Spearman}$ & +0.685 & +0.731 & -0.562 & +0.439 & +0.595 & -0.478 \\
    \noalign{\smallskip}
    \hline 
	\end{tabular}
	\end{table*}	
	
    \subsection{XO-2S}
    	\label{sec:stellaractivity2}
    By looking at the average value of the activity index \textit{R$^{\prime}_{\rm HK}$} (Table \ref{table:actindtable}), this star appears less active than its companion. In Fig. \ref{fig:rhktimexo2s} the time series of \textit{R$^{\prime}_{\rm HK}$} is shown. The internal errors are quite comparable to the data scatter, with no clear evidence of a long-term trend. 
    
    In the same way as for XO-2N, we investigated the relations between the asymmetry indicators and the \textit{R$^{\prime}_{\rm HK}$} index, and the results are shown in Fig. \ref{fig:asyindS}. No significant positive or negative correlation is found, supporting the scenario of a very quiet star. This outcome is confirmed by the results, shown in Table \ref{table:signcorrcoeffS}, of the analysis performed to asses the significance of the Spearman's rank correlation coefficients, as done for XO-2N. For completeness, we also quantified the correlation between the \textit{R$^{\prime}_{\rm HK}$} index and the RV residuals (after the subtraction of the two-planet Keplerian solution) calculated by \cite{desidera14}, obtaining a Spearman's rank coefficient $\rho$=0.27. By testing the null hypothesis, we found a z-score = -0.94 that corresponds to a significant probability p=17.4$\%$ that the two datasets are not correlated. This result is confirmed by applying the composite method, from which we obtain  $\rho=0.20\pm0.15$, indicating lack of correlation.
    
    We also used the GLS algorithm to search for periodicities in the CCF asymmetry indexes and \textit{R$^{\prime}_{\rm HK}$} time series, but no significant peaks were detected. For instance, for the BIS and \textit{$\Delta$V} parameters we found peaks close to half the tentative stellar rotation period derived from photometry (P$\sim$26 days). A bootstrap analysis resulted in FAP of 0.12 and 0.03, suggesting the the signals are spurious. No peaks are found close to the orbital periods of the two planets orbiting XO-2S.

	\begin{table}
	\scriptsize
	\centering
	\caption{Results of the analysis carried out to asses the significance of the Spearman's rank correlations coefficients for the pairs of XO-2S datasets shown in Fig. \ref{fig:asyindS}.}   
	 \label{table:signcorrcoeffS}
	\begin{tabular}{cccc}
	\hline
    \noalign{\smallskip}
     	Method	& R$^{\prime}_{\rm HK}$ \textit{vs} BIS & R$^{\prime}_{\rm HK}$ \textit{vs} $\Delta$V & R$^{\prime}_{\rm HK}$ \textit{vs} V$_{asy}$\\ 
      \noalign{\smallskip}
      \hline
      \noalign{\smallskip}	 	
    \textit{Null hypoth.} &  z-score = 0.45 & z-score = 0.37 & z-score = 0.69 \\ [3pt]
    & p-value = 32.6$\%$ &	 p-value = 47.2$\%$ &	p-value = 24.5$\%$ \\ [5pt]
    \textit{Composite MC} &  $\rho$ = -0.046$\pm$0.150  & $\rho$ = -0.008$\pm$0.145 & $\rho$ = -0.047$\pm$0.145  \\ 
    \noalign{\smallskip}
    \hline 
    \noalign{\smallskip}
    Original $\rho_{Spearman}$ & -0.068 & -0.054 & -0.101 \\
    \noalign{\smallskip}
    \hline 
	\end{tabular}
	\end{table}

    \subsection{Comparison of the activity levels for the XO-2 stars}
    	\label{sec:stellaractivity3}
    In Table \ref{table:actindtable} we summarize our activity-related measurements for the XO-2 components. We can conclude that, at least in the timespan covered by our observations, the star XO-2N appears to be slightly more active than the companion. This is also directly visible when comparing the \ion{Ca}{ii} H$\&$K lines of the coadded HARPS-N spectra (Fig. \ref{fig:coaddhk}), which show an emission flux in the line cores greater for the North component. Also, the differences seen in the existing correlations between the CCF asymmetry indicators and the activity index \textit{R$^{\prime}_{\rm HK}$} support the scenario that the activity level of the North component is higher. This appears in agreement with the higher levels of magnetic activity in stars with hot Jupiters with respect to their wide binary companions, as found in particular for HD 189733 and Corot-2 (see, e.g., \citealt{lanza11,husnoo12,poppen13,poppen14}). Then, the different activity levels of the two stars might be due to any influence of the hot Jupiter on the activity of XO-2N or to the fact that we simply observed the two stars at different epochs of their activity cycles.

    In Table \ref{table:actindtable} we also show the expected values of the spin rotation periods derived from the measurements $\log$(\textit{R$^{\prime}_{\rm HK}$}) index according to the works of \cite{noyes84} and \cite{mamajek08}. It can be seen that the estimate of the rotation period of XO-2N derived from photometry is in very good agreement with the expected values.


\section{Considerations about the long-term stability and evolution of the XO-2 planetary systems}
\label{sec:sysstab}

In a recent work, \cite{wang14} determined the planet occurrence rate in multiple-star systems by measuring the stellar multiplicity in a sample of \textit{Kepler} planet host stars, and comparing the stellar multiplicity rate between these and field stars in the solar neighbourhood. Interestingly, they find that planet formation appears to be significantly suppressed in multiple-star systems when the separation between the planet host star and the other stellar companion is smaller than 1500 AU, and this distance could be thus assumed as the radius of influence of stellar companions. Nonetheless, the estimate of the planet occurrence rate in multiple-star systems is presently a subject of debate. \cite{horch14} investigated the binary fraction of {\it Kepler} exoplanet host stars by observing more than 600 objects of interest, and conclude that this appears similar to the rate observed for field stars (40-50$\%$), pointing out that the impact of binarity on the planet occurrence should be lower than that emerging from the work of \cite{wang14}. 

The analysis of the stellar properties of the XO-2 binary points to the scenario that the components have a common origin and formed a bound system since their birth. Although the parameters of the binary orbit are not known, we have performed a basic analysis of the long-term stability of the whole system based on the work of \cite{holman99}, extending to the XO-2S system similar considerations done by \cite{burke07} for XO-2N. \cite{holman99} modelled the influence of a stellar companion on the stability of a 'test particle' orbiting a star with an empirical relation (see Eq. (1) therein) among the stellar mass ratio $\mu$=m$_{\rm 2}$/(m$_{\rm 1}$+m$_{\rm 2}$) (the subscript \textit{1} identifies the star orbited by a planet), the binary eccentricity \textit{e}, and the semi-major axis of the binary. This relation provides an estimate of the \textit{critical} semi-major axis \textit{a$_{\rm c}$}, i.e. the maximum orbital distance of a planet to survive after 10,000 binary orbits (corresponding to more than 10$^{9}$ yr in the case of XO-2). 

\begin{figure}
   \centering
   \includegraphics[width=9cm]{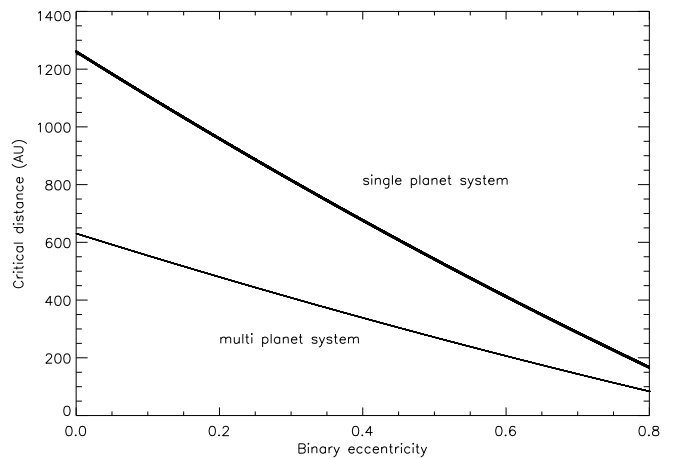}
   \caption{Critical semi-major axis \textit{a$_{\rm c}$} for each star in the XO-2 binary system, defined as the maximum distance that a single planet, on initially circular orbits, could have from its parent star to survive after 10,000 binary orbits (thick line). The thin line represents a simple estimate of \textit{a$_{\rm c}$} when one of the stars hosts more than one planet. These results are based on the work of \cite{holman99}.}
   \label{fig:critdist}%
    \end{figure}

The work of \cite{holman99} considers planets that orbit only one component of the stellar system, on initially circular and prograde orbits in the
plane of the binary. We can apply their relation to each XO-2 stars separately, obtaining similar results as the mass ratio is $\mu\sim$0.5 in both cases. By assuming the measured projected binary separation \textit{a}=4600 AU as an estimate of the actual semi-major axis of the orbit, and with the eccentricity of the binary orbit left undetermined, we can estimate the critical distance in the range of validity of the empirical formula, 0$\leqslant$\textit{e}$\leqslant$0.8. In the case of a single planet orbiting one of the components the dependence of the critical semi-major axis \textit{a$_{\rm c}$} from the eccentricity is shown in Fig. \ref{fig:critdist} by the thick curve. Following the discussion in Sect.5 of \cite{holman99}, we can assume half of the \textit{a$_{\rm c}$} values found in the case of the three-body problem as a rough estimate of the critical distance when one star hosts more planets, which is here the case of -at least- the star XO-2S. This result is shown by the thin line in Fig. \ref{fig:critdist}. According to it, if the binary has the highest eccentricity considered here (\textit{e}=0.8), a planetary companion around XO-2S would be in a stable orbit if located at a distance $\leqslant$90 AU from the star, while for XO-2N the upper limit is $\sim$170 AU. 

A final consideration should be made about the possible role of the Lidov-Kozai (L-K) mechanism (e.g. \citealp{innanen97, takeda08}) in particular on the migration of the planet XO-2Nb so close to its parent star. Because the orbital parameters of the binary are not known, at this stage we can not reconstruct the evolutionary history of this quite old system. In particular, we do not have any direct measurement of the mutual inclination between them and the binary orbital plane, from which the effects of L-K mechanism strongly depends. \cite{neveu14} discussed that the L-K mechanism likely played a role in shaping the orbits of the two hot Jupiter systems discovered around the components of the WASP-94 binary, this scenario being reinforced by the high projected spin-orbit angle \textit{$\lambda$}=151$^{\circ}\pm$20$^{\circ}$ and the retrograde orbit of WASP-94Ab. Any discussion about the evolution of the XO-2 planetary systems still remains mostly at the level of speculation. However, in our opinion by assuming that the orbital configuration of the binary did not change greatly with time, the L-K mechanism alone can not explain why the XO-2Nb migrated close to its parent star and became a hot Jupiter, while XO-2Sb and XO-2Sc are farther out, maybe being prevented to migrate because of mutual secular perturbations. In fact, the large separation of the two stars leads to very long periods of the L-K cycle, well beyond the present age of the system. Nonetheless, if the system originated closer to the galactic bulge, as discussed in Sect.\ref{sec:stellarpar3}, this could have produced a strong modification of its orbit with time, making the study of the perturbing mechanisms very complicated.
Despite our present inability in assessing the role of the L-K mechanism for XO-2 on firm basis, nonetheless it is interesting to mention the relevant work of \cite{jensen14} about the determination of the protoplanetary disk geometry in a binary system. Thanks to observations with the Atacama Large Millimeter Array (ALMA), they obtained the clearest picture ever seen of protoplanetary discs in a young binary (age < 4 Myr), located roughly at the same distance of the XO-2 system. They provided evidence for a relative inclination of the discs in the range 60$^{\circ}$-68$^{\circ}$, revealing that both or at least one of them must be misaligned with the binary orbital plane, and probably forming an angle far greater than the critical value of $\sim$40$^{\circ}$ for which the L-K mechanism is expected to be effective. The results of \cite{jensen14} suggest that the formation processes of a binary can favour those necessary conditions to perturb at some extent the planetary orbits since the first stages of planet formation.  

\section{Conclusions}

We have investigated in detail the physical properties of the XO-2 stellar and planetary systems. This is the first confirmed case of a wide binary whose components, with a T$_{\rm eff}$ similar to our Sun but older and super metal-rich, host planets. This circumstance motivated a first comprehensive characterization study. Our analysis is based on high-resolution spectroscopic data collected with the HARPS-N spectrograph, and new photometric data from several facilities. The main results described in this paper can be summarized as follows:

  \begin{enumerate}
     \item Through a spectroscopic analysis carried out with both the equivalent widths and spectral synthesis methods, and the use of stellar evolutionary tracks, we derived new estimates of the physical parameters of the XO-2 stars. By performing a differential spectral analysis,
we measured a positive difference in iron abundance of XO-2N with respect to XO-2S at more than $3\sigma$. When considering this finding in conjunction with the existence of a hot Jupiter around the star with the higher iron abundance, a very interesting implication is that the evolution of this planetary system, assuming the migration of the giant planet, could have favoured the ingestion by the host star of rocky material in the planetary disk, which can be revealed as an excess of iron abundance.
     \item Fourteen new transit light curves were collected of the hot Jupiter XO-2Nb. From their combined analysis we derived updated and very accurate transit parameters. 
     By analysing the out-of-transit light curve of XO-2N collected during a timespan of $\sim$4 months, we measured for the first time its rotation period. We found that the light curve can be folded almost perfectly with a sine function of period 41.4$\pm$1,1 days and that our estimate is well in agreement with predictions. We also provide an estimate for the rotation period of XO-2S, but our result is inconclusive because the interpretation of the periodogram is more complicated than in the case of XO-2N. Indeed, the highest peak corresponds to a rotation period of 26 days, which is $\sim$15~days shorter than that of XO-2N and far from being in agreement with gyrochronology predictions. However, another peak with almost comparable power is found at 34.5 days and, because of aliasing effects, for the moment the true periodicity can not be determined.
     \item We analysed the radial velocity time series of the XO-2N star, improving the estimates of the orbital parameters of XO-2Nb. As a major result, we confirmed the existence of an acceleration previously reported in the literature and revealed a turn-over in the trend. We discussed the implications of this finding in terms of the presence of an outer companion for which we determined a lower limit for its mass and orbital period. Nonetheless, we found a significant positive correlation between the residuals of the RV and the activity index \textit{R$^{\prime}_{\rm HK}$} which would imply that the observed acceleration is rather caused by the intrinsic activity of the star and related to its solar-like activity cycle. However, despite we found relevant and convincing indications for a stellar activity cycle as the real cause of the observed long-term RV variation, more observations would be advisable to definitively clarify the origin of the RV trend. Finally, by combining the results from the analysis of the transit light curves and of the radial velocity time series (archival data and new HARPS-N measurements), we refined the main physical parameters of the planet XO-2Nb, in particular the mass, radius and density. 
     \item We modelled the Rossiter-McLaughlin effect for XO-2N, providing a better constrain on the projected spin-orbit angle \textit{$\lambda$}=7$^{\circ}\pm$11$^{\circ}$, according to which the system is likely aligned. By combining the result on  \textit{$\lambda$} with that on the stellar rotation period, we could derive for the first time an upper limit of 39$^{\circ}$ for the true angle \textit{$\psi$} formed by the stellar rotation axis and the normal to the planetary orbital plane. 
     \item We investigated the stellar activity of the XO-2 components by analysing a set of asymmetry indicators of the spectral cross-correlation function and the chromospheric activity index \textit{R$^{\prime}_{\rm HK}$}. We conclude that the North component was more active during the timespan of our observations, and the different activity levels might be due to any influence of the hot Jupiter XO-2Nb on its host star or to different phases of their activity cycles.
  \end{enumerate}
     
  The XO-2 system should be still regarded with interest, in particular for the evidence of long-term trends in the RVs of both the stars for which a clear explanation is still not possible. Should these signals be caused by additional companions in outer orbits, this would make the system still more interesting to be studied as a peculiar laboratory for testing the theories of planet formation and evolution. We are collecting new data to improve the characterization of the whole system and provide insights to several open questions.


\begin{acknowledgements}
We thank the anonymous referee for relevant comments which improved the quality of the data analysis. The GAPS project acknowledges support from INAF through the “Progetti Premiali” funding scheme of the Italian Ministry of Education, University, and Research. The Astronomical Observatory of the Autonomous Region of the Aosta Valley is supported by the Regional Government of the Aosta Valley, the Town Municipality of Nus and the Mont Emilius Community. MD acknowledges partial support from INAF-OATo through the grant "Progetto GAPS: caratterizzazione spettroscopica e fotometrica dei target (attività cromosferica, rotazione) e studio delle sinergie tra GAPS e APACHE" ($\#$35/2014), and from ASI under contract to INAF I/058/10/0 (Gaia Mission – The Italian Participation to DPAC). JMC and AB are supported by a grant of the European Union-European Social Fund, the Autonomous Region of the Aosta Valley and the Italian Ministry of Labour and Social Policy. We thank ASI (through contracts I/037/08/0 and I/058/10/0) and Fondazione CRT for their support to the APACHE Project. VN acknowledges partial support from INAF-OAPd through the grant "Analysis of HARPS-N data in the framework of GAPS project" ($\#$19/2013) and "Studio preparatorio per le osservazioni della missione ESA/CHEOPS" ($\#$42/2013).
NCS acknowledges support from Funda\c{c}\~ao para a Ci\^encia e a Tecnologia (FCT, Portugal) through FEDER funds in program COMPETE, as well as through national funds, in the form of grants reference RECI/FIS-AST/0176/2012 (FCOMP-01-0124-FEDER-027493), and RECI/FIS-AST/0163/2012 (FCOMP-01-0124-FEDER-027492), and in the form of the Investigador FCT contract reference IF/00169/2012 and POPH/FSE (EC) by FEDER funding through the program "Programa Operacional de Factores de Competitividade - COMPETE. NCS further acknowledges the support from the European Research Council/European Community under the FP7 through Starting Grant agreement number 239953. 
IR acknowledges support from the Spanish Ministry of Economy and Competitiveness (MINECO) and the ”Fondo Europeo de Desarrollo Regional” (FEDER) through grants AYA2012-39612-C03-01 and ESP2013-48391-C4-1-R. We thank M. Fiaschi and A. Zurlo for their support during the TASTE observations of 2011/02/06 and 2011/04/02, respectively.
This research has made use of the SIMBAD database and VizieR catalogue access tool, operated at CDS, Strasbourg, France.
\end{acknowledgements}


\bibliographystyle{aa} 
\bibliography{ref.bib} 

\end{document}